\documentclass[fleqn,twoside]{article}
\usepackage{espcrc2}

\usepackage{graphicx}
\usepackage[figuresright]{rotating}

\newcommand{\AmS}{{\protect\the\textfont2
  A\kern-.1667em\lower.5ex\hbox{M}\kern-.125emS}}

\newcommand{\be}{\begin{equation}}
\newcommand{\ee}{\end{equation}}
\newcommand{\bea}{\begin{eqnarray}}
\newcommand{\eea}{\end{eqnarray}}
\newcommand{\ba}{\begin{eqnarray}}
\newcommand{\ea}{\end{eqnarray}}

\newcommand{\ihsp}{\hspace*{\fill} }

\newcommand{\Fig}[1]{Fig.~{\ref{#1}}}

\title{QCD modeling of hadron physics}

\author{P. Maris\address{Dept. of Physics and Astronomy, 
 University of Pittsburgh, Pittsburgh, PA 15260}%
\address[kent]{Center for Nuclear Research, 
 Dept. of Physics, Kent State University, Kent OH 44242}%
        and P.C. Tandy\addressmark[kent]%
}
       
\begin{document}

\begin{abstract}
We review recent developments in the understanding of meson properties
as solutions of the Bethe--Salpeter equation in rainbow-ladder
truncation.  Included are recent results for the pseudoscalar and
vector meson masses and leptonic decay constants, ranging from pions
up to $c\bar{c}$ bound states; extrapolation to $b\bar{b}$ states is
explored.  We also present a new and improved calculation of
$F_\pi(Q^2)$ and an analysis of the $\pi\gamma\gamma$ transition form
factor for both $\pi(140)$ and $\pi(1330)$.  Lattice-QCD results for
propagators and the quark-gluon vertex are analyzed, and the effects
of quark-gluon vertex dressing and the three-gluon coupling upon meson
masses are considered.
\vspace{1pc}
\end{abstract}

\maketitle

\section{DYSON--SCHWINGER EQUATIONS OF QCD}

The Dyson--Schwinger equations [DSEs] are the equations of motion of a
quantum field theory.  They form an infinite hierarchy of coupled
integral equations for the Green's functions ($n$-point functions) of
the theory.  Bound states (mesons, baryons) appear as poles in the
Green's functions.  Thus, a study of the poles in $n$-point functions
using the set of DSEs will tell us something about hadrons.  For
recent reviews on the DSEs and their use in hadron physics, see
Refs.~\cite{review,Alkofer:2000wg,Maris:2003vk}.

\subsection{Quark propagator}
The exact DSE for the quark propagator is\footnote{We use
Euclidean metric $\{\gamma_\mu,\gamma_\nu\} = 2\delta_{\mu\nu}$,
$\gamma_\mu^\dagger = \gamma_\mu$ and $a\cdot b = \sum_{i=1}^4 a_i
b_i$.}
\begin{eqnarray}
\lefteqn{S(p)^{-1}  =  i \not\!p\, Z_2+ m_q(\mu)\,Z_4 }
\nonumber \\ && {} + Z_1 \int_k \, g^2D_{\mu \nu}(q)
        \, \gamma_\mu {\hbox{$\frac{\lambda^i}{2}$}}
        \, S(k) \, \Gamma^i_\nu(k,p)  \,,
\label{Eq:quarkDSE}
\end{eqnarray}
where $D_{\mu\nu}(q=k-p)$ is the renormalized dressed gluon
propagator, and $\Gamma^i_\nu(k,p)$ is the renormalized dressed
quark-gluon vertex.  The notation $\int_k$ stands for $\int^\Lambda
d^4 k/(2\pi)^4$.  For divergent integrals a translationally invariant
regularization is necessary; the regularization scale $\Lambda$ is to
be removed at the end of all calculations, after renormalization, and
will be suppressed henceforth.

The solution of Eq.~(\ref{Eq:quarkDSE}) can be written as
\begin{eqnarray}
 S(p)  &=&  \frac{Z(p^2)}{i \not\!p + M(p^2)} \,,
\label{Eq:quarkprop}
\end{eqnarray}
renormalized according to $S(p)^{-1} = i\,/\!\!\!p + m(\mu)$ at a
sufficiently large spacelike $\mu^2$, with $m(\mu)$ the current quark
mass at the scale $\mu$.  Both the propagator, $S(p)$, and the vertex,
$\Gamma^i_\mu$ depend on the quark flavor, although we have not
indicated this explicitly.  The renormalization constants $Z_2$ and
$Z_4$ depend on the renormalization point and on the regularization
mass-scale, but not on flavor: in our analysis we employ a
flavor-independent renormalization scheme.

\subsection{Mesons}
Bound states correspond to poles in $n$-point functions: for example a
meson appears as a pole in the 2-quark, 2-antiquark Green's function
$G^{(4)}=\langle 0 | q_1 q_2 \bar{q_1} \bar{q_2} | 0 \rangle$.  In the
vicinity of a meson, i.e. in the neighborhood of $P^2 = -M^2$ with $M$
being the meson mass, such a Green's function behaves like
\begin{eqnarray}
 G^{(4)} &\sim& 
 \frac{\chi(p_o, p_i;P) \bar{\chi}(k_i,k_o;P)}{P^2 + M^2} \,,
\end{eqnarray}
where $P$ is the total 4-momentum of the meson, $p_o$ and $p_i$ are
the momenta of the outgoing quark and incoming quark respectively, and
similarly for $k_i$ and $k_o$.  Momentum conservation relates these
momenta: $p_o-p_i = P = k_o - k_i$.

The function $\chi(p_o,p_i;P)$ describes the projection of the bound
state onto its $q \bar{q}$ Fock-space component\footnote{Here and
subsequently, $q$ refers to dressed quark (quasi-particle) states.}
(sometimes referred to as the Bethe-Salpeter [BS] wavefunction, which
is different from the light-cone wave function).  It satisfies the
homogeneous Bethe--Salpeter equation [BSE]
\begin{eqnarray}
 \Gamma(p_o, p_i;P) = \int_k \! K(p_o,p_i;k_o,k_i) \,
	\chi(k_o, k_i;P) \, ,
\label{Eq:genBSE}
\end{eqnarray}
with $\Gamma$ the Bethe--Salpeter amplitude [BSA]
\begin{eqnarray}
\Gamma(k_o,k_i; P) = S(k_o)^{-1} \, \chi(k_o,k_i; P) \, S(k_i)^{-1} \,.
\end{eqnarray}
The kernel $K$ is the $q \bar{q}$ irreducible quark-antiquark
scattering kernel.  This homogeneous integral equation has solutions
at discrete values $P^2 = -M^2$ of the total meson 4-momentum $P$.

In order to identify the meson mass $M$ from the BSE,
Eq.~(\ref{Eq:genBSE}), it is convenient to first
introduce the linear eigenvalue $\lambda(P^2)$ of the kernel
\begin{eqnarray}
\lefteqn{ \Gamma(p_o, p_i;P) =}
\nonumber \\ && \lambda(P^2) \int_k\; K(p_o,p_i;k_o,k_i) \;
	\chi(k_o, k_i;P) \, ,
\label{Eq:BSEev}
\end{eqnarray}
which has solutions for all values of $P^2$.  Since we use a Euclidean
metric, we have to analytically continue this eigenvalue equation to
the timelike region, i.e. negative values of $P^2$, and then find $M$
such that \mbox{$\lambda(-M^2) = 1$}.  The corresponding eigenvector
$\Gamma(p_o,p_i;P)$ is then a solution of the original BSE,
Eq.~(\ref{Eq:genBSE}), i.e. it is the BSA of a $q\bar{q}$ bound state.
This meson BSA is normalized according to
\begin{eqnarray}
 \lefteqn{2\, P_\mu = N_c \; \frac{\partial}{\partial Q_\mu} \bigg\{ }
\nonumber\\ && \int_{k,q} {\rm Tr}\big[
          \bar\chi(k_i,k_o)\,
	  K(\tilde{k_o},\tilde{k_i};\tilde{q_o},\tilde{q_i})\,
          \chi(q_o,q_i) \big] + {}
\nonumber\\ && {} \int_k {\rm Tr}\big[
        \bar\Gamma(k_i,k_o)\,S(\tilde{k_o})\,
        \Gamma(k_o,k_i)\,S(\tilde{k_i}) \big]
\bigg\} \bigg|_{Q=P}
\nonumber\\ {}
\label{Eq:BSEnorm}
\end{eqnarray}
at $P^2=-M^2$, with $k_o-k_i=P=q_o-q_i$ and 
$\tilde{k_o}-\tilde{k_o}=Q=\tilde{q_o}-\tilde{q_i}$.

The properly normalized BSA $\Gamma(p_o,p_i;P)$ [or equivalently,
$\chi(p_o,p_i;P)$] completely describes the meson as a $q\bar{q}$
bound state.  Different types of mesons, such as pseudoscalar or
vector mesons, are characterized by different Dirac structures.  The
ground state in any particular spin-flavor channel corresponds to the
largest eigenvalue $\lambda$ in that channel being one.  The first
excited state is determined by the next-largest eigenvalue $\lambda$
being one, and so on for higher excited states.

\section{RAINBOW-LADDER TRUNCATION}

A viable truncation of the infinite set of DSEs has to respect
relevant (global) symmetries of QCD such as chiral symmetry, Lorentz
invariance, and renormalization group invariance.  For electromagnetic
interactions we also need to respect current conservation.  The
so-called rainbow-ladder truncation respects these properties.  In
this truncation, the kernel of the meson BSE is replaced by an
(effective) one-gluon exchange
\begin{eqnarray}
\lefteqn{ K(p_o,p_i;k_o,k_i) \to }
\nonumber \\ &&
        -4\pi\,\alpha(q^2)\, D_{\mu\nu}^{\rm free}(q)
        \textstyle{\frac{\lambda^i}{2}}\gamma_\mu \otimes
        \textstyle{\frac{\lambda^i}{2}}\gamma_\nu \,,
\end{eqnarray}
where $q=p_o-k_o=p_i-k_i$, and $\alpha(q^2)$ is an effective running
coupling.  The corresponding truncation of the quark DSE is
\begin{eqnarray}
\lefteqn{
Z_1 g^2 D_{\mu \nu}(q) \Gamma^i_\nu(k,p) \to }
\nonumber \\ &&
 4\pi\,\alpha(q^2) \, D_{\mu\nu}^{\rm free}(q)\, \gamma_\nu
                                        \textstyle\frac{\lambda^i}{2} \,.
\label{Eq:rainbow}
\end{eqnarray}
This truncation is the first term in a systematic
expansion~\cite{systematicexp} of the quark-antiquark scattering kernel
$K$; asymptotically, it reduces to leading-order perturbation theory.
Furthermore, these two truncations are mutually consistent in the
sense that the combination produces vector and axial-vector vertices
satisfying their respective Ward identities.  In the axial case, this
ensures that in the chiral limit the ground state pseudoscalar mesons
are the massless Goldstone bosons associated with chiral symmetry
breaking~\cite{Maris:1997hd,Maris:1997tm}.  In the vector case, this
ensures, in combination with impulse approximation, electromagnetic
current conservation~\cite{Roberts:1994hh}.  We will come back to this
point later, when we discuss electromagnetic processes.

\begin{figure}[tb]
\includegraphics[width=7.cm]{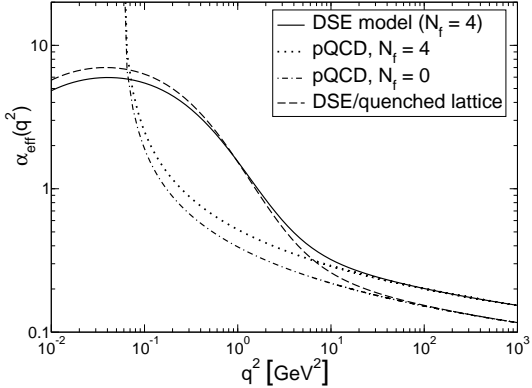}
\caption{
The effective ladder-rainbow kernel from the Maris--Tandy [MT]
model~\protect\cite{Maris:1999nt} compared to LO perturbative QCD and
to that deduced from quenched lattice QCD data on the gluon and quark
propagators~\protect\cite{Bhagwat:2003vw}.
\label{Fig:MT_kernel}}
\end{figure}
The ultraviolet behavior of the effective running coupling is dictated
by the one-loop renormalization group equation; the infrared behavior
of the effective interaction is modeled, and constrained by
phenomenology.  Here, we are using a 2-parameter model for the
effective interaction~\cite{Maris:1999nt}, fitted to give a value for
the chiral condensate of $(240~{\rm MeV})^3$ and $f_\pi=131~{\rm
MeV}$.  This effective running coupling, with $\Lambda_{\rm QCD} =
0.234~{\rm GeV}$ and $N_f = 4$, is shown in Fig.~\ref{Fig:MT_kernel},
and agrees perfectly with pQCD for $q^2 > 25~{\rm GeV}^2$.  The model
is finite in the infrared region, and it is very similar to an
effective interaction that was recently deduced~\cite{Bhagwat:2003vw}
from quenched lattice QCD results for the gluon and quark propagators.
Indications are that unquenching does not change the infrared behavior
of the running coupling much~\cite{Fischer:2003rp}, and thus we
believe that this 2-parameter model is a realistic parametrization of
the effective quark-quark interaction in the spacelike region.

\subsection{Quark propagator}
A momentum-dependent quark mass function $M(p^2)$ is an essential
property of QCD.  In the perturbative region this mass function reduces to
the one-loop perturbative running quark mass
\begin{eqnarray}
  M(p^2) &\simeq& \frac{\hat{m}}
  {\left(\frac{1}{2}\ln\left[p^2
   /\Lambda_{\rm QCD}^2\right]\right)^{\gamma_m}} \,,
\label{eq:currentM}
\end{eqnarray}
with $\gamma_m = 12/(33-2N_f)$ the anomalous mass dimension.
Dynamical chiral symmetry breaking [D$\chi$SB] means that this mass
function is nonzero even when the current quark mass is zero.  In the
chiral limit the mass function is~\cite{Politzer:1976tv}
\begin{eqnarray}
\lefteqn{ M_{\hbox{\scriptsize{chiral}}}(p^2) \, \simeq }
\nonumber \\ &&  \frac{2\pi^2\gamma_m}{3}\,
        \frac{-\,\langle \bar q q \rangle^0}{p^2 \left(
        \frac{1}{2}\ln\left[p^2/\Lambda_{\rm QCD}^2\right]
    \right)^{1-\gamma_m}}\,,
\label{eq:chiralM}
\end{eqnarray}
with $\langle \bar q q \rangle^0$ the
renormalization-point-independent vacuum quark
condensate~\cite{Maris:1997tm}.

\begin{figure}[tb]
\includegraphics[width=7.cm]{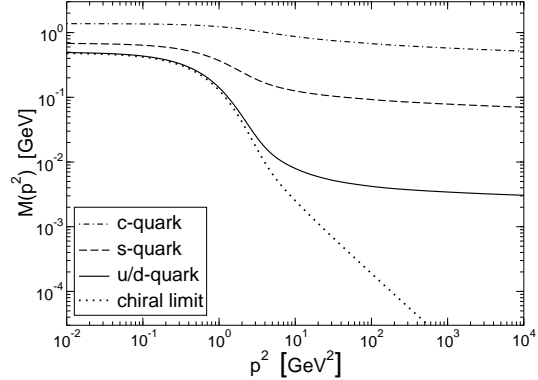}
\caption{Dynamical quark mass function using the rainbow-ladder
truncation of Ref.~\cite{Maris:1999nt}.
\label{Fig:quarkprp}}
\end{figure}
It is a longstanding prediction of DSE studies in QCD that both the
mass function $M(p^2)$ and the wavefunction renormalisation function
$Z(p^2)$ receive strong momentum-dependent corrections at infrared
momenta, see e.g. Refs.~\cite{review,Alkofer:2000wg,Maris:2003vk} and
references therein.  Provided that the (effective) quark-quark
interaction reduces to the perturbative running coupling in the
ultraviolet region, it is also straightforward to reproduce the
asymptotic behavior of Eqs.~(\ref{eq:currentM}) and
(\ref{eq:chiralM})~\cite{Higashijima:1983gx,Fomin:1984tv}.  Both these
phenomena are illustrated in Fig.~\ref{Fig:quarkprp}, using the model
for the interaction of Fig.~\ref{Fig:MT_kernel}.  Here we can also see
that in the infrared region the dynamical mass function of the $u$ and
$d$ quarks becomes very similar to that of chiral quarks.  This is a
direct consequence of D$\chi$SB, and leads to a dynamical mass
function of several hundred MeV for the light quarks in the infrared
region, even though the corresponding current quark masses are only a
few MeV.  Thus we connect the current quark mass of perturbative QCD
with a ``constituent-like'' quark mass at low energies.

These predictions for the quark mass function have been confirmed in
lattice simulations of
QCD~\cite{Skullerud:2001aw,Bowman:2002kn,Bowman:2004xi}.  Pointwise
agreement for a range of quark masses requires this interaction to be
flavor-dependent~\cite{Bhagwat:2003vw}, suggesting that dressing the
quark-gluon vertex $\Gamma^i_\nu(q,p)$ is important.  The consequences
of a dressed vertex for the meson BSEs are currently being
explored~\cite{Bhagwat:2004hn}.  This will be discussed in more detail
in Sec.~\ref{Sec:beyond}.

\subsection{Pseudoscalar and vector mesons}
The most general structure of a pseudoscalar meson BSA 
is
\begin{eqnarray}
\lefteqn{ \Gamma^{\rm PS}(k+\eta P,k-(1-\eta)P) = }
\nonumber \\
        && \gamma_5 \big[ i E + \;/\!\!\!\! P \, F + 
        \,/\!\!\!k \, G + \sigma_{\mu\nu}\,k_\mu P_\nu \,H \big]\,,
\label{Eq:decompPS}
\end{eqnarray}
where the invariant amplitudes $E$, $F$, $G$ and $H$ are Lorentz
scalar functions of $k^2$ and $k\cdot P$, and they also depend on the
choice for $\eta$.  The natural choice for mesons with equal-mass
constituents (like a pion) is $\eta=\frac{1}{2}$, though physical
observables are of course independent of $\eta$.

In the chiral limit, the axial-vector Ward--Takahashi identity [WTI]
relates the pseudoscalar meson BSA to the (inverse) quark
propagator~\cite{Maris:1997hd}.  As a consequence, if there is
D$\chi$SB, there is a massless pseudoscalar bound state (Goldstone
boson) and vice versa: a massless pseudoscalar bound state indicates
that chiral symmetry is broken dynamically.  Furthermore, we find
\begin{eqnarray}
  E(p^2,p^2) &=& \frac{M(p^2)}{f_\pi \; Z(p^2)} \,.
\end{eqnarray}
Thus, in the chiral limit we do not have to solve any bound state
equation in order to obtain the coefficient function of the canonical
Dirac structure of a pseudoscalar meson.  This also shows why
constituent quark models are inadequate in describing pions: for a
realistic description of pions as $q\bar{q}$ bound states one
necessarily has to have a (strongly) momentum dependent quark mass
function.

Away from the chiral limit, the axial-vector WTI relates the
pseudoscalar and pseudovector spin projections of the BS wavefunction
\begin{eqnarray}
 f_{\rm PS} \, M_{\rm PS}^2 &=& 2\, m_q(\mu) \, r_{\rm PS}(\mu) \,,
\label{Eq:GMOR}
\end{eqnarray}
where
\begin{eqnarray}
  f_{\rm PS} \, P_\mu &=& Z_2 \, N_c \int_k 
{\rm Tr}[\chi(k_o, k_i) \, \gamma_5 \, \gamma_\mu] \, ,
\\
  r_{\rm PS}(\mu) &=& -i Z_4 \, N_c \int_k 
{\rm Tr}[\chi(k_o, k_i) \, \gamma_5] \,.
\end{eqnarray}
\begin{figure}[tb]
\includegraphics[width=7.5cm]{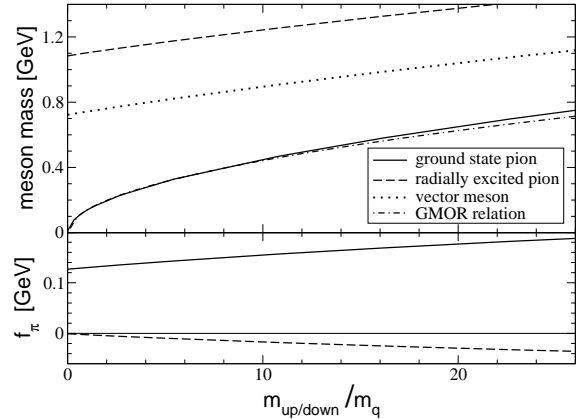}
\caption{Pion and excited pion mass and decay constant, as function
of the current quark mass (normalized by the realistic $u/d$ quark masses);
for comparison we have also included the vector meson mass.
\label{Fig:cairnsGMOR}}
\end{figure}
This is an exact relation in QCD, and holds independent of the current
quark mass.  The Gell-Mann--Oakes--Renner relation is a direct
corollary of this relation, but that is not the only observable
consequence.  It also means that the decay constant $f_{\rm PS}$ of
excited pions vanishes in the chiral limit~\cite{Holl:2004fr}. This
relation is indeed satisfied in the rainbow-ladder truncation, as can
be seen from Fig.~\ref{Fig:cairnsGMOR}, both for the ground state pion
and for its first radial excitation~\cite{Holl:2004fr}.  Furthermore,
for heavy-light pseudoscalar mesons it predicts~\cite{Ivanov:1997iu}
that the decay constant decreases like $1/\sqrt{M}$.

The next-lightest mesons are the vector mesons.  The BSA of a massive
vector meson can be decomposed into eight Dirac structures
\begin{eqnarray}
\lefteqn{ \Gamma^{\rm V}_\mu(k+\eta P,k-(1-\eta)P) = }
\nonumber \\
        && \sum_{i=1}^8 f^i(k^2, k\cdot P; \eta) \, T^i_\mu(k,P) \,,
\label{Eq:decompV}
\end{eqnarray}
where $T^i_\mu(k,P)$ are eight independent transverse Dirac tensors.
Again, the invariant amplitudes $f^i$ are Lorentz scalar functions of
$k^2$ and $k\cdot P$, and depend on $\eta$.

\begin{table}[bt]
\caption{
DSE results~\cite{Maris:1999nt} for the meson masses and decay
constant, together with experimental data~\cite{PDG}. 
\label{Table:model} }
\renewcommand{\tabcolsep}{1pc} 
\renewcommand{\arraystretch}{1.2} 
\begin{tabular}{@{}lcc}
\hline
        & experiment   & calculated        \\
        & (estimates)  & ($^\dagger$ fitted) \\ \hline
$m^{u=d}_{\mu=1 {\rm GeV}}$ &
        \multicolumn{1}{r}{ 4 - 11 MeV}  &
        \multicolumn{1}{r}{ 5.5 MeV}     \\
$m^{s}_{\mu=1 {\rm GeV}}$ &
        \multicolumn{1}{r}{ 100 - 300 MeV} &
        \multicolumn{1}{r}{ 125 MeV   }    \\ \hline
- $\langle \bar q q \rangle^0_{\mu}$
                & (0.24 GeV)$^3$ & (0.241$^\dagger$)$^3$ \\
$m_\pi$         &  0.138 GeV &    0.138$^\dagger$ \\
$f_\pi$         &  0.131 GeV &    0.131$^\dagger$  \\
$m_K$           &  0.496 GeV  &   0.497$^\dagger$ \\
$f_K$           &  0.160 GeV  &   0.155        \\ \hline
$m_\rho$        &  0.770 GeV  &   0.742        \\
$f_\rho$        &  0.216 GeV  &   0.207        \\
$m_{K^\star}$   &  0.892 GeV  &   0.936        \\
$f_{K^\star}$   &  0.225 GeV  &   0.241        \\
$m_\phi$        &  1.020 GeV  &   1.072        \\
$f_\phi$        &  0.236 GeV  &   0.259        \\ \hline
\end{tabular}\\[2pt]
\end{table}
With the model of Ref.~\cite{Maris:1999nt}, we obtain good agreement
with the experimental values for the light pseudoscalar and vector
meson masses and leptonic decay constants, see
Table~\ref{Table:model}.  These results show little sensitivity to
variations in the model parameters, as long as the integrated strength
of the effective interaction is strong enough to generate an
acceptable amount of chiral symmetry breaking, as indicated by the
chiral condensate.  This is not true for heavier states: e.g. the
radially excited pion is quite sensitive to details of the
interaction~\cite{Holl:2005vu}.

\subsection{Frame independence}
The BSE is usually solved in the rest-frame of the meson.  To be
explicit, the most convenient frame is that characterized by
$P_\mu=(i\,M, 0, 0, 0)$.  However, this is not required: One of the
advantages of the DSE approach to hadron physics is its explicit
Poincar\'e covariance.  This makes it a particularly useful tool for
studying electromagnetic form factors and other processes.  No matter
what frame one chooses for calculating say the pion electromagnetic
form factor, at least one of the pions is moving.

\begin{figure}[h]
\includegraphics[width=7.cm]{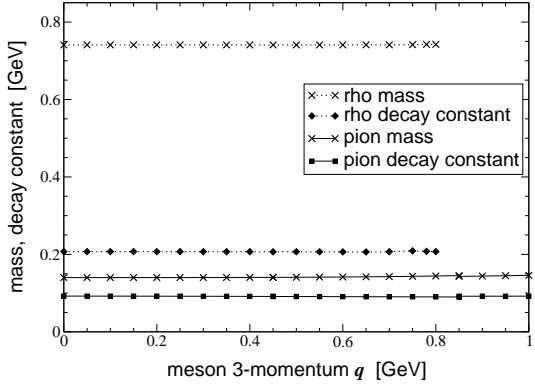}
\caption{Pion and $\rho$ mass and decay constant calculated in a
moving frame, as function of the meson 3-momentum.
\label{Fig:framedep}}
\end{figure}

As a test, we have calculated the static $\pi$ and $\rho$ properties
in a moving frame $P_\mu = (i\,E, q, 0, 0)$ where $q$ is the
3-momentum of the moving meson.  Within this frame we solve again the
eigenvalue equation, Eq.~(\ref{Eq:BSEev}), search for $P^2 = -M^2$
such that $\lambda(P^2) = 1$, and calculate the corresponding
electroweak decay constant.  Numerically, this is a bit of a
tour-de-force, since the Lorentz scalar functions of
Eqs.~(\ref{Eq:decompPS}) and (\ref{Eq:decompV}) are now functions of a
radial variable $k^2$ and {\em two} angles
\begin{eqnarray}
 k\cdot P &=& i\, k\,E \cos\alpha + k\,q\, \sin\alpha \cos\beta \,,
\end{eqnarray}
and the integral equation has to be solved in the three independent
variables $k^2$, $\alpha$, and $\beta$.  With current computer
resources, this can be done without further approximations, and the
results, shown in Fig.~\ref{Fig:framedep}, are indeed independent of
the meson 3-momentum, illustrating that this approach is indeed
Poincar\'e covariant.  We can now use this same approach to calculate
meson form factors in an explicitly covariant manner.

\section{ELECTROMAGNETIC PROCESSES}

In order to study electromagnetic processes, we need to know the
dressed quark-photon coupling, in addition to the meson BSAs and the
dressed quark propagators.

\subsection{Quark-photon vertex}
The $q\bar{q}\gamma$ vertex is the solution of the renormalized
inhomogeneous BSE with the same kernel $K$ as the homogeneous BSE for
meson bound states.  That is, for photon momentum $Q$, we have
\begin{eqnarray}
\Gamma_\mu(p_o,p_i) &=& Z_2\, \gamma_\mu + \int_k  K(p_o,p_i;k_o,k_i)
\nonumber \\ && {}\times
         S(k_o)\,\Gamma_\mu(k_o,k_i) \, S(k_i) \,,
\label{Eq:vectorBSE}
\end{eqnarray}
with $p_o$ and $p_i$ the outgoing and incoming quark momenta,
respectively, and similarly for $k_o$ and $k_i$, with
$p_o-p_i=k_o-k_i=Q$.  Because of gauge invariance, the vertex
satisfies the vector WTI
\begin{equation}
i\,Q_\mu \,\Gamma_{\mu}(p_+,p_-)  =
                       S^{-1}(p_+) - S^{-1}(p_-) \,.
\label{Eq:vecWTI}
\end{equation}

The most general form of the quark-photon vertex
$\Gamma_\mu(p_o,p_i;Q)$ requires a decomposition into twelve Dirac
structures.  Four of these covariants represent the longitudinal
components which are completely specified by the WTI in terms of the
(inverse) quark propagator and they do not contribute to elastic form
factors.  The transverse vertex can be decomposed into eight Dirac
tensors $T^i_\mu(p;Q)$ with the corresponding amplitudes being Lorentz
scalar functions.

Note that solutions of the {\em homogeneous} version of
Eq.~(\ref{Eq:vectorBSE}) define vector meson bound states with masses
\mbox{$M_{\rm V}^2=-Q^2$} at discrete timelike momenta $Q^2$.  It
follows that $\Gamma_\mu$ has poles at those locations and, in the
neighborhood of $Q^2 = -M_{\rm V}^2$, behaves like~\cite{Maris:1999bh}
\begin{eqnarray}
 \Gamma_\mu(p_o,p_i) &\sim& 
 \frac{\Gamma_\mu^{\rm V}(p_o,p_i) \, f_{\rm V} \, M_{\rm V}}
      {Q^2 + M_{\rm V}^2} \; ,
\end{eqnarray}
where $\Gamma_\mu^{\rm V}$ is the vector meson BSA, and $f_V$ the
electroweak decay constant.  The fact that the dressed
$q\bar{q}\gamma$ vertex exhibits these vector meson poles lies behind
the success of naive vector-meson-dominance [VMD] models; the effects
of intermediate vector meson states on electromagnetic processes can
be unambiguously incorporated by using the properly dressed
$q\bar{q}\gamma$ vertex rather than the bare vertex
$\gamma_\mu$~\cite{Maris:1999bh}.

\subsection{Impulse approximation}
The generalized impulse approximation allows electromagnetic processes
to be described in terms of dressed quark propagators, bound state
BSAs, and the dressed $q\bar{q}\gamma$-vertex.  Consider for example
the 3-point function describing the coupling of a photon with momentum
$Q$ to a pseudoscalar meson $a\bar{b}$, with initial and final momenta
\mbox{$P \pm Q/2$}, which can be written as the sum of two terms
\begin{eqnarray}
\Lambda^{a\bar{b}}_\mu(P,Q) &=&
                \hat{Q}^a \, \Lambda^{aa\bar{b}}_\mu
                + \hat{Q}^{\bar{b}} \, \Lambda^{a\bar{b}\bar{b}}_\mu \,,
\end{eqnarray}
where $\hat{Q}$ is the quark or antiquark electric charge, and where
$\Lambda^{a\bar{b}a}(P,Q)$ and $\Lambda^{a\bar{b}\bar{b}}(P,Q)$
describe the coupling of a photon to the quark ($a$) and antiquark
($\bar{b}$) respectively.  In impulse approximation, these couplings
are given by
\begin{eqnarray}
\Lambda^{aa\bar{b}}_\mu(P,Q) &=&
        i\,N_c \int_k \!{\rm Tr}\big[ \Gamma^{a}_\mu(q_-,q_+)
	  \, \chi^{a\bar{b}}(q_+,q) 
\nonumber \\ && {} \times
        S^b(q)^{-1} \, \bar\chi^{\bar{b}a}(q,q_-)\big] \;,
\label{Eq:triangle}
\end{eqnarray}
with \mbox{$q = k-P/2$} and \mbox{$q_\pm = k+P/2 \pm Q/2$}, and
similarly for $\Lambda^{a\bar{b}\bar{b}}_\nu$.  The corresponding
meson elastic form factor is defined by
\begin{equation}
 \Lambda^{a\bar{b}}_\mu(P,Q) = 2\;P_\mu\;F(Q^2) \,.
\end{equation}

\subsection{Current conservation}
Electromagnetic current conservation dictates $F(0) = \hat{Q}^a +
\hat{Q}^{\bar{b}}$ and $Q_\mu \Lambda^{a\bar{b}}_\mu(P,Q) = 0$.  It is
straightforward to show that both these conditions are satisfied in
impulse approximation, in combination with rainbow-ladder truncation.
At \mbox{$Q=0$} the $q\bar{q}\gamma$ vertex is completely specified by
the differential Ward identity
\begin{equation}
i \,\Gamma_{\mu}(p,p) =
        \frac{\partial}{\partial p_\mu} S^{-1}(p) \; .
\label{Eq:diffWI}
\end{equation}
If this is inserted in Eq.~(\ref{Eq:triangle}), one finds after a
change of integration variables
\begin{eqnarray}
\lefteqn{ \Lambda^{aa\bar{b}}_\mu(P,0) =
        2 \, P_\mu \, F^{aa\bar{b}}(0) = }
\nonumber \\  &&
        N_c \int_q \! {\rm Tr}\bigg[
        \Gamma^{a\bar{b}}(q',q) \, S^b(q) \,
        \bar{\Gamma}^{\bar{b}a}(q,q') \,
        \frac{\partial S^a(q')}{\partial P_\mu} \bigg]\,,
\nonumber \\ {}
\end{eqnarray}
with $q'=q+P$.  Comparing this expression to Eq.~(\ref{Eq:BSEnorm}),
we recognize that the requirement \mbox{$F^{aa\bar{b}}(0)=1$} follows
directly from the canonical normalization condition for
$\Gamma^{a\bar{b}}$ provided that (1) the BSE kernel $K$ is
independent of the total meson momentum $P$ and (2) the
$a\bar{a}\gamma$ vertex satisfies the differential Ward
identity~\cite{Roberts:1994hh}.

The second constraint, $Q_\mu \Lambda^{a\bar{b}}_\mu(P,Q) = 0$, is
satisfied if the quark-photon vertex satisfies the vector WTI.
Inserting Eq.~(\ref{Eq:vecWTI}) into Eq.~(\ref{Eq:triangle}), we find
\begin{eqnarray}
\lefteqn{ Q_\mu \, \Lambda^{aa\bar{b}}_\mu(P,Q) =
        N_c \int_k \!{\rm Tr}\big[ 
 \chi^{a\bar{b}}(q_+,q) \, \bar\Gamma^{\bar{b}a}(q,q_-) \big] }
\nonumber \\ && {}
 - N_c \int_k \!{\rm Tr}\big[ 
 \Gamma^{a\bar{b}}(q_+,q) \, \bar\chi^{\bar{b}a}(q,q_-) \big] \;,
\end{eqnarray}
which clearly vanishes.

The ladder BSE kernel is indeed independent of the meson momentum $P$,
and with quark propagators dressed in rainbow approximation and the
dressed quark-photon vertex calculated in ladder approximation, the
both the vector WTI and the differential Ward identity are
satisfied~\cite{Maris:1999bh}; thus impulse approximation for the
quark-photon vertex in combination with rainbow-ladder truncation is
consistent in the sense that the correct electric charge of the meson
is produced independent of the model parameters and the resulting
meson electromagnetic current is conserved.

\subsection{Pion form factor}
\begin{figure}[tb]
\includegraphics[width=7.5cm]{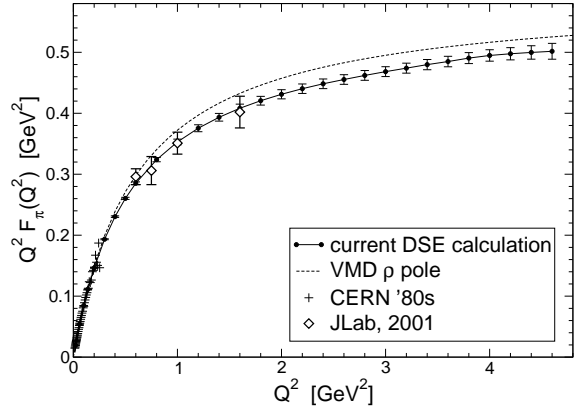}
\caption{
Our results for $Q^2 \, F_\pi(Q^2)$, with estimates of our numerical
error bars, compared to a VMD model and to experimental data from
CERN~\cite{Amendolia:1986wj} and JLab~\cite{Volmer:2000ek}.
\label{Fig:q2fpi}}
\end{figure}

For the pion electromagnetic form factor, working in the
isospin limit, we have
\begin{eqnarray}
2\,P_\mu\,F_\pi(Q^2) &=&
        i\,N_c \int_k \!{\rm Tr}\big[ \chi^\pi(q,q_+)
        \, \Gamma_\nu(q_+,q_-)
\nonumber \\ && {} \times
        \bar\chi^\pi(q_-,q)\,S(q)^{-1} \big] \;.
\label{Eq:pionEMF}
\end{eqnarray}
with the propagators solutions of the quark DSE in rainbow truncation.
The pion BSAs and the $q\bar{q}\gamma$ vertex are solutions of their
respective BSEs in ladder truncation, using exactly the same momentum
frame as used in the form factor calculation.  That does mean
re-calculating the pion BSAs for every value of the photon $Q^2$, but
the advantage is that one does not have to do any interpolation or
extrapolation on the numerical solutions of the BSE.  Our results for
spacelike $Q^2$ are shown in Fig.~\ref{Fig:q2fpi}.  They are in
excellent agreement with the available experimental
data~\cite{Amendolia:1986wj,Volmer:2000ek} and with our previous
results\footnote{The current calculations improve on
Refs.~\cite{Maris:1999bh,Maris:2000sk} by eliminating: the Chebyshev
expansion for the angular dependence of the restframe BSE solutions,
and the interpolation and extrapolation techniques used to connect
frames.}.

Technical difficulties due to the appearance of a pair of complex
conjugate singularities in the quark propagator limit our calculation
to $Q^2 < 5~{\rm GeV}^2$.  Up to this value of $Q^2$ we see no
significant deviation from a naive monopole behavior in our
calculation of the form factor.  It would be very interesting to see
whether or not TJNAF can establish a clear deviation from a monopole
behavior, because pQCD predicts a significantly lower value for $Q^2
F_\pi(Q^2)$ at (asymptotically) large $Q^2$.

\subsection{Quark mass dependence}
We have calculated $F_\pi(Q^2)$ for a range of current quark masses.
For all masses considered, the form factor behaves like a monopole,
with a monopole mass that is slightly smaller than the $\rho$-meson
mass.  A similar behavior has been found in quenched lattice
simulations~\cite{vanderHeide:2003kh,Bonnet:2004fr}.

\begin{figure}[tb]
\includegraphics[width=7.cm]{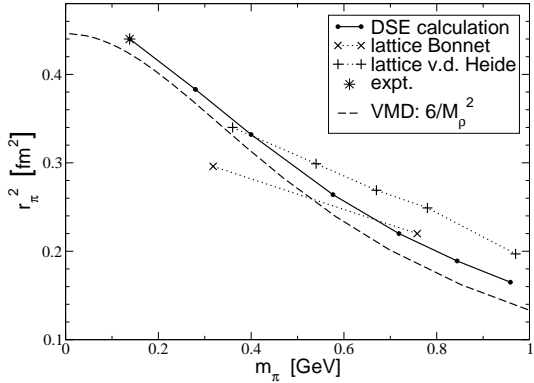}
\caption{
Pion charge radius as function of the quark mass within our DSE model,
and as extracted from lattice
simulation~\cite{vanderHeide:2003kh,Bonnet:2004fr}.
\label{Fig:rpi}}
\end{figure}
The obtained charge radius, defined by \mbox{$r^2 = -6 F'(0)$}, is
also in excellent agreement with the experimental value.  Our
calculated charge radius as a function of the quark mass is shown in
Fig.~\ref{Fig:rpi}, together with the charge radius obtained from
lattice simulations.  For comparison, we also show the charge radius
one would obtain from a naive VMD model, using the corresponding
(calculated) vector meson mass.  Again, both our numerical result for
$r_\pi^2$ and that of lattice simulations are represented reasonably
well by a VMD model for the charge radius.

An effect that is missing both from our calculations and from quenched
lattice simulations are pion loop corrections.  At present it is not
clear how to incorporate meson loops self-consistently in both the
DSE/BSEs and in the approximation for the photon-hadron coupling.
However, in Ref.~\cite{Alkofer:1993gu} it was demonstrated that the
dressed quark core generates most of the pion charge radius, and that
pion loops contribute less than 15\% to $r_\pi^2$ at the physical
value of the pion mass.  For larger values of the current quark mass
(and thus of the pion mass), we expect these corrections to become
negligible.  Also for larger spacelike values of $Q^2$ the effects
from meson loops decrease, and for \mbox{$Q^2 > 1 \,{\rm GeV}^2$} we
expect the contribution of such loops to be negligible.

\subsection{$\pi\,\gamma\,\gamma$ transition form factor}
Using impulse approximation, the coupling of a pion to two photons is,
in the isospin limit
\begin{eqnarray} 
\lefteqn{\Lambda^{\pi\gamma\gamma}_{\mu\nu}(Q_1, Q_2) = 
\alpha_{\rm em}\frac{4\pi\,\sqrt{2}\, N_c}{3} }
\nonumber \\ && 
\int_k \!{\rm Tr}\big[ \chi^\pi(q,p) 
\Gamma^\gamma_\mu(p,k)\, S(k) \, \Gamma^\gamma_\nu(k,q) \big] \;,
\\
\nonumber \\ & = & \frac{2\,i\,\alpha_{\rm em}}{\pi} 
\,\epsilon_{\mu \nu \alpha \beta }\,Q_{1\alpha }Q_{2\beta }
        G_{\pi\gamma\gamma}(Q_1^2,Q_2^2)  \,,
\end{eqnarray}
where the internal momenta $p, q$ are determined in terms of the
integration momentum $k$ and the external momenta $Q_1, Q_2$ by
momentum conservation; furthermore, the pion is on-shell, but the
photons not necessarily.  

For on-shell photons, $Q_1^2 = 0 = Q_2^2$, the partial decay width
$\pi^0 \to \gamma \gamma$ is given by
\begin{eqnarray}
 \Gamma_{\pi \gamma \gamma} &=&
	\frac{\alpha^2_{\rm em} \, m_\pi^3}{16 \pi^3 }\,
	G_{\pi\gamma\gamma}^2(0,0) \; .
\end{eqnarray}
This decay width is governed by the axial anomaly as prescribed by
electromagnetic current conservations and chiral symmetry.  In the
chiral limit, the axial anomaly gives\footnote{This expression is
conventionally given in terms of \mbox{$\hat{f}_\pi = f_\pi
/\sqrt{2}$}, the pion decay constant in the convention where its value
is $92$ MeV, rather than the convention \mbox{$f_\pi=131$} MeV used
throughout this work.}
\begin{equation}
G_{\pi\gamma\gamma}(0,0) = \frac{1}{2\,\hat{f}_\pi }
 \approx 5.4~{\rm GeV}^{-1} \,,
\label{Eq:anomaly}
\end{equation}
and the resulting decay width
\begin{eqnarray}
 \Gamma_{\pi \gamma \gamma} &=&
	\frac{\alpha^2_{\rm em} \, m_\pi^3}
	{64 \pi^3 \hat{f}_\pi^2} \; ,
\end{eqnarray}
is within 2\% of the experimental width of $7.8~{\rm eV}$.  The 
corrections due to finite pion mass are small.

\begin{figure}[tb]
\includegraphics[width=7.cm]{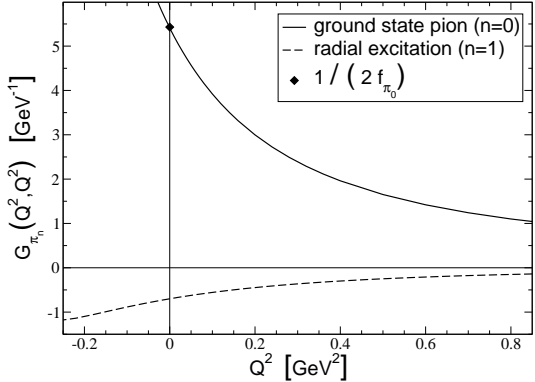}
\caption{ 
Symmetric $\pi\gamma\gamma$ transition form factor $G(Q^2, Q^2)$ for
both the ground state pion and its first radial
excitation. Adapted from~\cite{Holl:2005vu}.
\label{Fig:piggsym}}
\end{figure}

The axial anomaly is preserved by the ladder-rainbow truncation of the
DSEs combined with an impulse approximation for the $\pi^0 \gamma
\gamma$ vertex because the relevant manifestations of electromagnetic
gauge invariance and chiral symmetry are
present~\cite{Roberts:1994hh,Maris:1998hc}.  We do indeed reproduce
the above value for this form factor, as can be seen from
Fig.~\ref{Fig:piggsym}. Our results also agree with the experimental
data for the transition form factor
$G_{\pi\gamma^\star\gamma}(Q^2,0)$~\cite{Maris:2000wz,Maris:2002mz}.

On the other hand, the radially excited pion effectively decouples
from the axial anomaly~\cite{Holl:2005vu}, and its coupling to two
on-shell photons is not constrained by the anomaly.  A naive
application of Eq.~(\ref{Eq:anomaly}) to the first radially excited
pion would suggest an enormously large value of $625~{\rm GeV}^{-1}$
for its coupling $G(0,0)$ to two photons, because its decay constant
is very small, $\hat{f}_{\pi_1} \approx -0.0016~{\rm GeV}$, and in the
chiral limit this decay constant vanishes, as is evident from
Eq.~(\ref{Eq:GMOR}) and Fig.~\ref{Fig:cairnsGMOR}.  However,
numerically we find that for the excited pion this coupling is about
$0.7~{\rm GeV}^{-1}$.  Furthermore, the chiral limit $\pi\gamma\gamma$
form factor is at low-$Q^2$ almost identical to the curves shown in
Fig.~\ref{Fig:piggsym}, both for the ground state and for the excited
pion.

The asymptotic behavior of this form factor is given by the lightcone
operator product expansion~\cite{asympigg}
\begin{eqnarray}
 G(Q_1^2,Q_2^2) &\rightarrow&
        2 \pi^2 \hat{f}_{\pi_n}\bigg\{\frac{J(\omega)}{Q_1^2+Q_2^2} 
\nonumber \\ && {} + 
        {\cal O}\left(\frac{\alpha_s}{\pi}, \frac{1}{(Q_1^2+Q_2^2)^2}
        \right) \bigg\}\,,
\label{eq:gFOPE}
\end{eqnarray}
where $\omega = (Q_1^2 - Q_2^2)/(Q_1^2 + Q_2^2)$ is the photon
asymmetry.  In lightfront QCD $J(\omega)$ is related to the leading
twist pion distribution amplitude $\phi_\pi(x)$ via
\begin{eqnarray}
  J(\omega) &=& \frac{4}{3}\int_0^1 \!\!
        \frac{dx}{1-\omega^2(2x - 1)} \phi_\pi(x) \,.
\end{eqnarray}
The normalization of $\phi_\pi(x)$ immediately gives $J(0) =
\frac{4}{3}$ for the case of equal photon virtuality.  

\begin{figure}[tb]
\includegraphics[width=7.5cm]{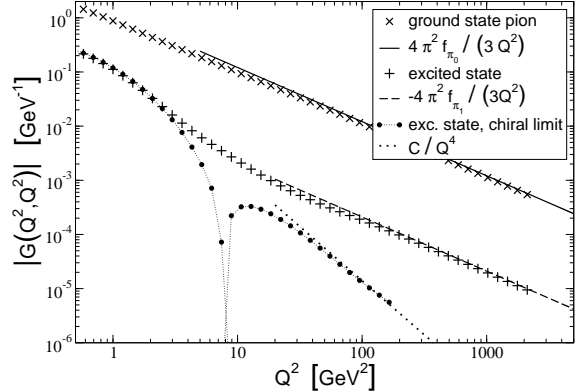}
\caption{Asymptotic behavior of the symmetric $\pi\gamma\gamma$
transition form factor. Adapted from~\cite{Holl:2005vu}.
\label{Fig:piggsymlog}}
\end{figure}
Within the present DSE-based approach, it is straightforward to
reproduce this leading-order asymptotic behavior analytically, not
only for the ground state pion~\cite{Maris:2002mz}, but also for its
excited states~\cite{Holl:2005vu}.  Also numerically we reproduce
\begin{eqnarray}
 G(Q^2,Q^2) &\to & \frac{4 \pi^2 \hat{f}_{\pi_n}}{3\,Q^2} \,,
\label{Eq:piggas}
\end{eqnarray}
as can be seen from Fig.~\ref{Fig:piggsymlog}.

The interesting question is what happens with the form factor of the
(first) radially excited pion in the chiral limit.  With a nonzero
current quark mass, this form factor indeed behaves like
Eq.~(\ref{Eq:piggas}), and it approaches zero from below as $Q_1^2 =
Q_2^2 \to \infty$, since its decay constant is negative.  In the
chiral limit however, the decay constant $\hat{f}_{\pi_1}$ is zero, as
dictated by Eq.~(\ref{Eq:GMOR}), and evident in
Fig.~\ref{Fig:cairnsGMOR}.  It turns out that in that case the form
factor vanishes like $1/Q^4$, with a positive coefficient: the form
factor has a zero-crossing in the spacelike region.  The coefficient
for this $1/Q^4$ behavior depends on the details of the model and its
parameters, but it is not simply related to $\hat{f}_{\pi_1}$.  The
leading order behavior of Eq.~(\ref{Eq:piggas}) however is
model-independent.

\section{MORE MASSIVE MESONS}

The rainbow-ladder truncation of the set of DSEs, using the model
interaction of Ref.~\cite{Maris:1999nt}, has been very successful in
describing a wide range of light meson properties~\cite{Maris:2003vk}.
One question that naturally arises: How does it do for heavier mesons?

\subsection{Heavy quark mesons}
\begin{figure}[tb]
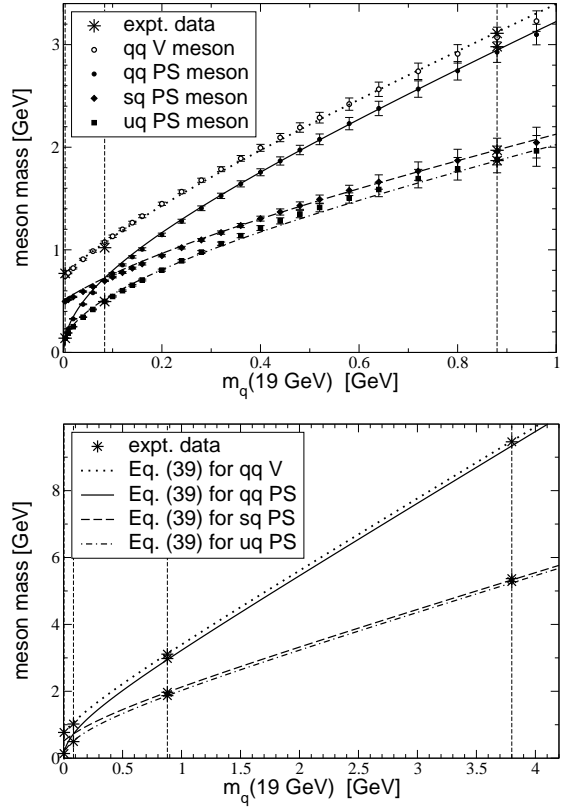

\includegraphics[width=7.3cm]{mesonmassfit.eps}

\vspace{7pt}
\includegraphics[width=7.3cm]{heavymass.eps}
\caption{ 
Meson mass as function of $m_q$.  Top: our calculation (with estimates
of numerical errors) up to $m_c$ plus the fit Eq.~(\ref{Eq:massfit});
bottom: extrapolated to $m_b$ using the same fit; the
vertical dashed lines indicate $m_{u/d}$, $m_s$, $m_c$, and $m_b$.
\label{Fig:massdep}}
\end{figure}
Let us first address the question of the quark mass dependence of the
pseudoscalar and vector meson properties.  In Fig.~\ref{Fig:massdep}
we show the meson mass and decay constants as function of $m_q(19)$,
the current quark mass at the renormalization point $\mu = 19~{\rm
GeV}$, both for heavy-light mesons (light meaning $u/d$ quarks and $s$
quarks) and for heavy-heavy mesons.

From this figure it is apparent that all of the meson masses scale
approximately linear with the current quark mass above about
$m_q\approx 0.5 ~{\rm GeV} \approx 6 m_s$, with $m_s$ being the
strange quark mass.  The pseudoscalar meson masses show a clear
curvature below this mass scale, as one would expect based on the
Gell-Mann--Oakes--Renner relation.  Also the vector meson masses show
a slight curvature at small current quark masses.  The results for the
$D$, $D_s$, $\eta_c$, and $J/\Psi$ mesons are in remarkably good
agreement with experiments: the calculated masses with $m_c(19) =
0.88~{\rm GeV}$ are within 2\% of the experimental values.  Compared
to the light current quark masses, we find $m_c/m_u = 238$ and
$m_c/m_s = 10.5$.

Both the vector and the pseudoscalar masses can be fitted reasonably
well over the entire range $0 < m_q < m_c$ by the same
formula
\begin{eqnarray}
  M^2 = M_0^2 + a_1 (m_1 + m_2) + a_2 (m_1 + m_2)^2 \,,
\label{Eq:massfit}
\end{eqnarray}
as indicated by the lines in Fig.~\ref{Fig:massdep}.  Here $m_i$ are
the current quark masses of the constituents (at $\mu=19~{\rm GeV}$);
the chiral limit masses are $M^{\rm PS}_0=0$, $M_0^V=0.75~{\rm GeV}$;
and the remaining fit parameters are $a_1^{\rm PS} = 2.96~{\rm GeV}$,
$a_1^{\rm V} = 3.24~{\rm GeV}$, and $a_2=1.12$ both for pseudoscalar
and vector mesons.

If we use this fit to extrapolate our results to the bottom quark
mass, we get the bottom panel of Fig.~\ref{Fig:massdep}.  The results
for the $B$, $B_s$, and $\Upsilon$ mesons are in surprisingly
agreement with experiments: the calculated masses with $m_b(19) =
3.8~{\rm GeV}$ are within 1\% of the experimental values.  Compared to
the other current quark masses, we find $m_b/m_c = 4.3$, $m_b/m_s =
45$ and $m_b/m_u$ is about 1,000.  Our prediction for $\eta_b$ is
$M_{\eta_b}=9.34~{\rm GeV}$

\begin{figure}[bt]
\includegraphics[width=7.3cm]{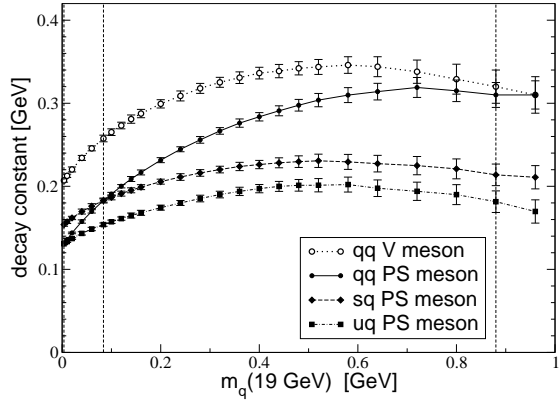}
\caption{ 
Meson decay constant (with estimates of our numerical errors) as
function of $m_q$, up to the charm quark mass region; the vertical
dashed lines indicate $m_{u/d}$, $m_s$, and $m_c$.
\label{Fig:mesondecay}}
\end{figure}

In Fig.~\ref{Fig:mesondecay} shows the corresponding leptonic decay
constants.  At small current quark masses, all of the decay constants
increase with the quark mass, but they tend to level off between the
$s$ and the $c$ quark mass, both for the heavy-light and for the
heavy-heavy mesons.  This is consistent with the expected behavior in
the heavy-quark limit: namely a decrease of the decay constant with
increasing meson mass (or equivalently, increasing heavy-quark mass)
like $f \sim 1/\sqrt{M}$, at least for the heavy-light mesons.  Our
current calculations suggest that the onset for this asymptotic
behavior could be as low as the charm quark mass for the $q\bar{u}$
mesons.  However, our results for both $f_D$ and $f_{D_s}$ are about
20\% below their experimental values, $f_D = 0.222 \pm 0.020~{\rm
GeV}$~\cite{Artuso:2005ym} and $f_{D_s} = 0.266 \pm 0.032~{\rm
GeV}$~\cite{PDG} respectively, and our result~\cite{Krassnigg:2004if}
for the decay constant of the $J/\psi$ is about 25\% below its
experimental value, extracted from the $e^+e^-$ decay~\cite{PDG}.
This could indicate that the rainbow-ladder truncation is not reliable
in the charm quark region, and/or our model for the effective
interaction is not applicable to charm quarks, despite the fact that
the meson masses are in good agreement with experiments.  Since the
decay constants depend on the norm of the BSAs, which in turn depends
on the derivative of the quark propagators, it is not a surprise that
the decay constants are more sensitive to details of the model than
the meson masses, and are thus better indicators for deficiencies in
the modeling.

\subsection{Heavier $u$ and $d$ quark mesons}
Another question is that of heavier states consisting of $u$ and $d$
quarks.  Calculations of the $a_1$ and $b_1$ mesons using a similar
model give results for these masses that are several hundred MeV too
low~\cite{Alkofer:2002bp,Watson:2004kd} when compared to experiments.
Also the mass of the first radially excited pion is, within this
model, significantly smaller than the experimental
value~\cite{Holl:2004fr}.  Furthermore, these heavier mesons, as well
as the scalar mesons~\cite{Maris:2002yu}, tend to be more sensitive to
details of the interaction than the $\rho$ and $\pi$.  

This suggests a deficiency of the model and/or of the rainbow-ladder
truncation for these higher-mass states.  A possible explanation for
this deficiency lies in the fact that these higher-mass states have a
different spin structure and (especially in case of the radially
excited pion) very different BS wave functions, and therefore they are
sensitive to different aspects of the BS kernel $K$.  Again, we should
study the effects from contributions beyond the rainbow-ladder
truncation in more detail in order to resolve these issues.

\section{BEYOND THE RAINBOW
\label{Sec:beyond}}

Both lattice simulations~\cite{Skullerud:2003qu,Skullerud:2004gp} and
DSE studies~\cite{Bhagwat:2004hn,Bhagwat:2004kj,Alkofer:2004it}
of the quark-gluon vertex indicate that $\Gamma^i_\nu(q,p)$ deviates
significantly from a bare vertex in the nonperturbative region.  A
(flavor-dependent) nonperturbative vertex dressing could make a
significant difference for the solution of the quark
DSE~\cite{Fischer:2003rp,Bhagwat:2004hn}.  Simple dressed vertex
models have indicated that material contributions to a number of
observables are possible with a better understanding of the infrared
structure of the vertex.  These diverse model indications include: an
enhancement in the quark
condensate~\cite{Fischer:2003rp,Bhagwat:2004hn}; an increase of about
300~MeV in the $b_1/h_1$ axial vector meson mass~\cite{Watson:2004kd};
and about 200~MeV of attraction in the $\rho/\omega$ vector meson
mass~\cite{Bhagwat:2004hn}.

\subsection{Lattice-inspired rainbow model}    
Over the last five years, the gluon 2-point function has become fairly
well known through explicit lattice-QCD
calculations~\cite{Leinweber:1998uu} and also through approximate
solutions of the gauge sector
DSEs~\cite{Alkofer:2000wg,Fischer:2003rp,vonSmekal:1997is}.  A strictly
defined rainbow truncation of the quark DSE for calculation of the
dressed quark propagator would proceed from this input.  However, the
modern information on the dressed gluon propagator shows a strong
infrared suppression and it is not possible to obtain a realistic
value for the condensate $\langle \bar{q} q \rangle$ in rainbow
truncation~\cite{Hawes:1998cw}.  In this sense it is clear that the
empirically successful rainbow-ladder kernels developed earlier
implicitly include effects of quark-gluon vertex dressing for the
quark DSE.  Although such an effective kernel shows a bare vertex
Dirac matrix structure, the infrared parameterization produces a
strength and dependence upon gluon momentum that is over and above
that of the gluon propagator.

\begin{figure}[bt]
\includegraphics[width=7.5cm]{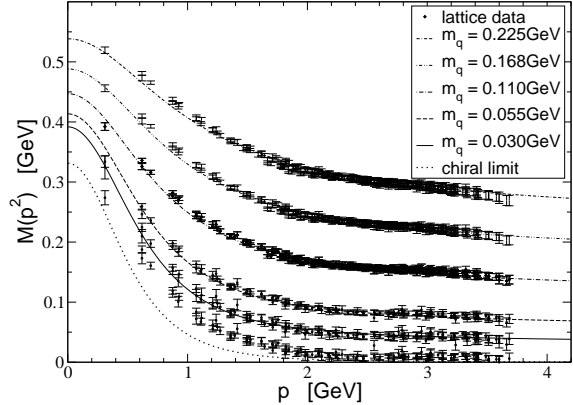}
\caption{ 
Dynamical quark mass functions obtained with the DSE-Lat model
together with the quenched lattice data~\cite{Bowman:2002kn}.
Adapted from~\protect\cite{Bhagwat:2003vw}.
\label{Fig:DSE_quLatt_fit}}
\end{figure}
This point is illustrated in
Fig.~\ref{Fig:DSE_quLatt_fit}~\cite{Bhagwat:2003vw}, where quenched
lattice data on the gluon and quark propagators are used to deduce a
phenomenological quark-gluon vertex by means of the DSE.  The DSE
kernel of Ref.~\cite{Bhagwat:2003vw} is modeled by
$4\pi\alpha(q^2)\,D_{\mu\nu}^{\rm free}(q) \to V(q^2, m_q)\,
D_{\mu\nu}^{\rm lat}(q)$ in Eq.~(\ref{Eq:rainbow}), where
$D_{\mu\nu}^{\rm lat}(q)$ represents the quenched lattice result for
the Landau gauge gluon propagator, and $V(q^2, m_q)$ is a
phenomenological representation of vertex dressing whose ultraviolet
behavior is determined by the requirement that the resulting running
coupling reproduces the one-loop renormalization group behavior of
QCD.  Thus, the vertex dressing has been mapped onto a single
amplitude corresponding to the canonical Dirac matrix $\gamma_\mu$,
and depends on the gluon momentum and the current quark mass.

The resulting effective DSE kernel [DSE-Lat] is compared to the MT
model in Fig.~\ref{Fig:MT_kernel}.  The parameters are determined by
requiring that the DSE solutions reproduce the quenched lattice
data~\cite{Bowman:2002kn} for $S(p)$ in the available domain
\mbox{$p^2 < 10~{\rm GeV}^2$} and \mbox{$m(\mu=2~{\rm GeV}) < 200~{\rm
MeV}$}.  In this sense, the DSE-Lat model represents quenched
dynamics.  It is found that the necessary vertex dressing is a strong
but finite enhancement.  As one would expect, the vertex dressing
$V(q^2, m_q)$ decreases with increasing $m_q$ to represent the effect
of quark propagators internal to the vertex.

The model easily reproduces $m_\pi$ with a current mass that is within
acceptible limits.  However the resulting chiral condensate
\mbox{$\langle \bar{q} q \rangle =(0.19~{\rm GeV})^3$} is a factor of 2
smaller than the empirical value $(0.24 \pm 0.01~{\rm GeV})^3$.  This
is attributed to the quenched approximation in the lattice data.

\subsection{Modeling the quark-gluon vertex}
\begin{figure}[tb]
\includegraphics[width=35mm]{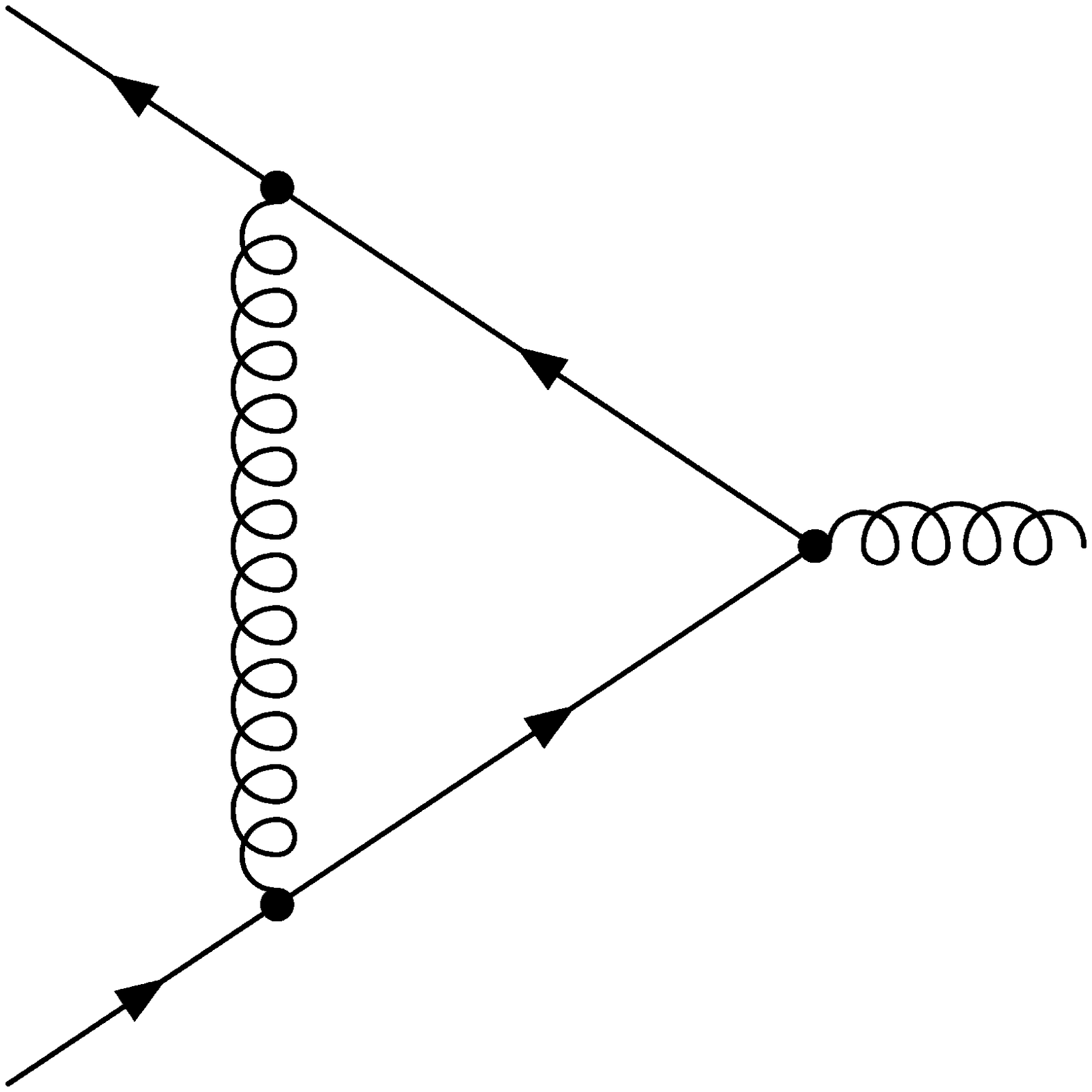}
\includegraphics[width=35mm]{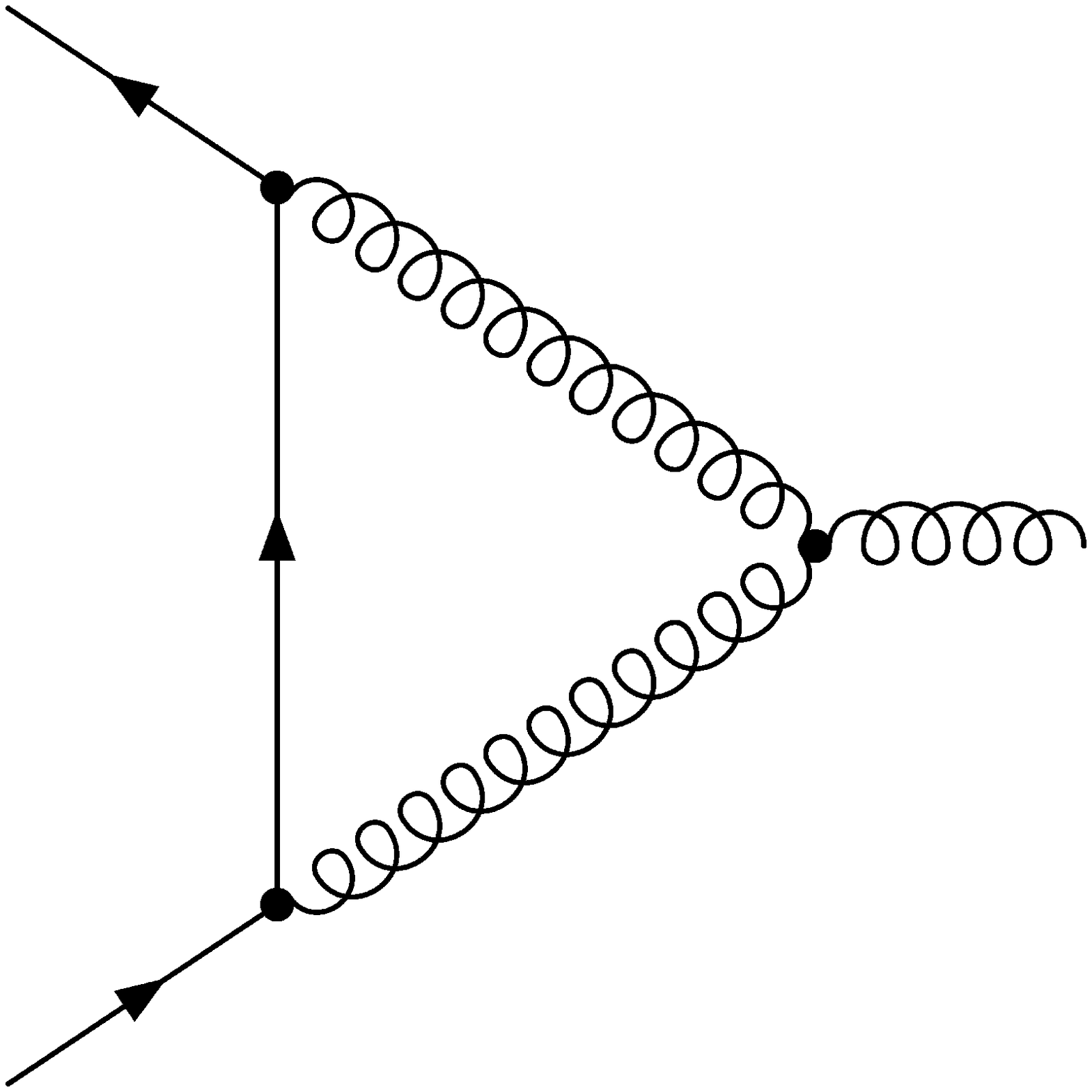}
\vspace*{-15mm}

\centerline{\ihsp A \ihsp \hspace*{20mm} NA \ihsp}

\caption{\label{fig:2vertdiags} The one-loop corrections to the 
quark-gluon vertex dressing. 
Left: the Abelian-like term $\Gamma_\sigma^{\rm A}$; Right:  
the non-Abelian term $\Gamma_\sigma^{\rm NA}$. }
\end{figure}

Let us denote the dressed quark-gluon vertex for gluon momentum $k$
and quark momenta $p$ and $p+k$ by
\mbox{$ig\,\frac{\lambda^i}{2}\,\Gamma_\sigma(p+k,p)$}.  Through
${\cal O}(g^2)$, i.e., to one loop, we have
\mbox{$\Gamma_\sigma=Z_{\rm 1F}\,\gamma_\sigma + \Gamma_\sigma^{\rm A}
+ \Gamma_\sigma^{\rm NA} $}, see \Fig{fig:2vertdiags}.  Here $Z_{\rm
1F}$ is the vertex renormalization constant to ensure
\mbox{$\Gamma_\sigma = \gamma_\sigma$} at the renormalization scale
$\mu$.

The color factor for $ \Gamma_\sigma^{\rm A}$ is \mbox{$(C_{\rm
F}-\frac{1}{2}C_{\rm A}) =-\frac{1}{2 N_c}$}, whereas the color factor
for $\Gamma_\sigma^{\rm NA}$ is \mbox{$\frac{1}{2}C_{\rm
A}=\frac{N_c}{2}$}.  This reveals two important issues: The
$\Gamma_\sigma^{\rm A}$ contribution to the quark-gluon vertex is
repulsive, whereas the non-Abelian $\Gamma_\sigma^{\rm NA}$ term,
which involves the three-gluon vertex, is attractive.  Furthermore,
the magnitude of the non-Abelian term $\Gamma_\sigma^{\rm NA}$ is
enhanced by a factor $N_c^2$ over the $\Gamma_\sigma^{\rm A}$
term\footnote{Note that the color factor of the color-singlet
$\Gamma_\sigma^{\rm A}$ contribution to the strong dressing of the
quark-photon vertex is \mbox{$C_{\rm F} = \frac{N_c^2-1}{2 N_c}$},
i.e. attractive and enhanced by $N_c^2$ over the color-octet
$\Gamma_\sigma^{\rm A}$ term: single gluon exchange between a quark
and antiquark has relatively weak repulsion in the color-octet
channel, compared to the strong attraction in the color-singlet
channel.}.

The term $\Gamma_\sigma^{\rm NA}$ involves the triple-gluon vertex
\mbox{$\Gamma^{3g}_{\mu \nu \sigma}$}, which is a ``new'' interaction.
However, in Landau gauge the corresponding (bare) Slavnov--Taylor
identity reduces $\Gamma_\sigma^{\rm NA}$ at zero gluon momentum,
$k=0$, to
\bea
\Gamma_\sigma^{\rm NA}(p,p) &=& - i\,\frac{C_{\rm A}}{2} \int_q\! 
\gamma_\mu S_0(p-q) \gamma_\nu\, \nonumber\\
&& \times \left\{\frac{\partial}{\partial q_\sigma} g^2\,D_0(q^2)
\right\}  T_{\mu \nu}(q) \,.
\label{VertNA0}
\eea

This expression has been used in a recent model~\cite{Bhagwat:2004kj}
for the dressed quark-gluon vertex at $k=0$.  It is a non-perturbative
extension of the two diagrams in \Fig{fig:2vertdiags}.  In the
ultraviolet, the $\bar q q$ scattering kernel appearing in
$\Gamma_\sigma^{\rm A}$ coincides with the ladder-rainbow kernel;
hence the natural extension is \mbox{$g^2\, D_0(q^2) \to 4 \pi
\alpha_{\rm eff}(q^2)/q^2$} and the external quark-gluon vertex is
taken to be bare.

For $\Gamma_\sigma^{\rm NA}$ it takes full advantage of the
simplification, Eq.~(\ref{VertNA0}), that happens for zero momentum
gluons: replace \mbox{$g^2\, D_0(q^2)$} by \mbox{$4 \pi \alpha_{\rm
eff}(q^2)/q^2$}.  This term is similar to a contribution to the
derivative of the quark self-energy but the differences are important;
at one-loop this term provides the explicitly non-Abelian
contributions to the Slavnov-Taylor identity~\cite{Davydychev:2000rt}.
The justifications for this nonperturbative vertex model are
consistency and simplicity; no new parameters are introduced.

\begin{figure}[t] 
\centerline{\includegraphics[width=7.5cm]{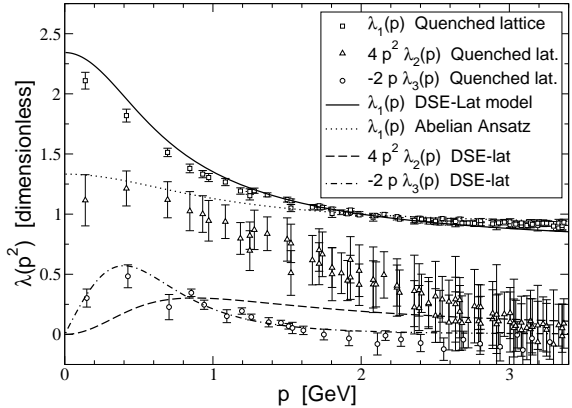}}
\caption{\label{fig:Lat_WI} 
The amplitudes of the dressed quark-gluon vertex at $k=0$ for
\mbox{$m_q(2~{\rm GeV}) = 60~{\rm MeV}$}.  Quenched lattice
data~\protect\cite{Skullerud:2003qu} are compared to the results of
the DSE-Lat model~\protect\cite{Bhagwat:2003vw}, and for
$\lambda_1(p)$ the Abelian Ansatz is also shown.
Adapted from~\cite{Bhagwat:2003vw}.}
\end{figure}

In \Fig{fig:Lat_WI} we display the results of this model in a
dimensionless form for comparison with the (quenched) lattice data for
\mbox{$m(\mu) = 60$}~MeV.  The general vertex at $k=0$ has a
representation in terms of four invariant amplitudes, but only 3 are
non-zero here.  We write
\ba
\Gamma_\sigma(p,p) &=& \gamma_\sigma \lambda_1(p^2)
     - 4 p_\sigma\, \gamma \cdot p\, \lambda_2(p^2) \nonumber \\
    &&  - i 2 p_\sigma\, \lambda_3(p^2)\,,
\label{lambda_def}
\ea 
since the lattice data~\cite{Skullerud:2003qu} are provided in terms
of these $\lambda_i(p^2)$ amplitudes.  The renormalization scale for
both the lattice data and the DSE-Lat model is $\mu = 2~{\rm GeV}$
where $\lambda_1(\mu) = 1$.  Without parameter adjustment, the model
reproduces the lattice data for $\lambda_1$ and $\lambda_3$ quite
well.  The lattice result for $\lambda_2$, despite the large errors,
suggests infrared strength that is seriously underestimated by the
DSE-Lat model.
  
The relative contributions to the vertex dressing made by $\Gamma^{\rm
NA}_\sigma$ and $\Gamma^{\rm A}_\sigma$ are indicated by the following
amplitude ratios at \mbox{$p=0$}: \mbox{$\lambda^{\rm
NA}_1/\lambda^{\rm A}_1 = -60$}, \mbox{$\lambda^{\rm
NA}_2/\lambda^{\rm A}_2 = -14$}, and \mbox{$\lambda^{\rm
NA}_3/\lambda^{\rm A}_3 = -12$}.  Thus the non-Abelian term
$\Gamma^{\rm NA}_\sigma$ dominates to a greater extent than what the
ratio of color factors ($-9$) would suggest.

Another useful comparison is the corresponding vertex in an Abelian
theory like QED; it is given by the differential Ward identity,
Eq.~(\ref{Eq:diffWI}).  With \mbox{$S^{-1}(p) =$} \mbox{$ i\gamma
\cdot p\, A(p^2) + B(p^2)$}, this leads to the correspondance
\mbox{$\lambda_1^{\rm WI} = A$}, \mbox{$\lambda_2^{\rm WI} =$} \mbox{$
-A^\prime/2$}, and \mbox{$\lambda_3^{\rm WI} =$} \mbox{$ B^\prime$},
where \mbox{$f^\prime =$} \mbox{$ \partial f(p^2)/ \partial p^2$}.
The Abelian Ansatz~\cite{Ball:1980ay}, while clearly inadequate for
the dominant amplitude $\lambda_1$ below 1.5~GeV, does reproduce the
DSE-Lat results for both $\lambda_2$ and $\lambda_3$.

\subsection{Effect on meson observables}
There is a known constructive scheme~\cite{systematicexp} that defines
a diagrammatic expansion of the BSE kernel corresponding to any
diagrammatic expansion of the quark self-energy such that the vector
and axial-vector WTIs are preserved.  Among other things, this
guarantees the Goldstone boson nature of the flavor non-singlet
pseudoscalars independently of model details~\cite{Maris:1997hd}.
\begin{figure}[ht] 
\includegraphics[width=7.cm]{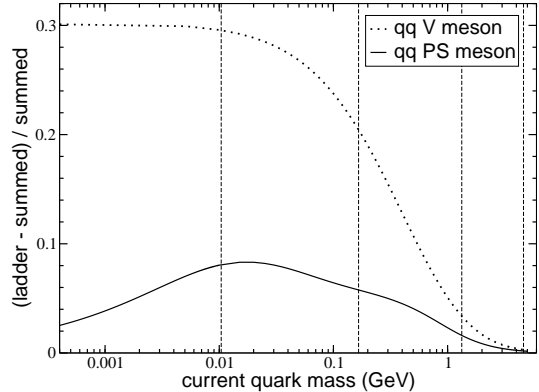}
\caption{\label{fig:LR_err} The relative error of the ladder-rainbow
truncation relative to the completely resummed model of vertex
dressing and corresponding BSE kernel. Adapted from
Ref.~\protect\cite{Bhagwat:2004hn}.  }
\end{figure} 

This has recently been exploited to produce the first
study~\cite{Bhagwat:2004hn} of how an infinite sub-class of
quark-gluon vertex diagrams, including attraction from three-gluon
coupling, contributes to vector and pseudoscalar meson masses.  The
symmetry-preserving BSE kernel obtained from the self-energy has an
infinite sub-class of diagrams beyond ladder; it is manageable here
because of the algebraic structure afforded from use of a modification
of the Munczek-Nemirovsky $\delta$-function
model~\cite{Munczek:1983dx}.  It is verified that the pseudoscalar
$q\bar{q}$ meson remains a Goldstone boson at any order of truncation.
As shown in \Fig{fig:LR_err}, the ladder-rainbow truncation has at
most a 8\% repulsive error as $m_q$ is varied up to the $b$ quark
region.

\Fig{fig:LR_err} also shows that attraction from the dressed vertex is
a significant correction to ladder-rainbow truncation for the
$q\bar{q}$ light quark vector mesons, and reduces steadily in relative
importance as $m_q$ increases.  This is consistent with the
results~\cite{Bhagwat:2003vw} shown in Fig.~\ref{Fig:DSE_quLatt_fit}
where the effective vertex dressing $V(q^2, m_q)$ required by quenched
lattice-QCD propagators also decreases with (relatively light) $m_q$.
It is also consistent with our results, shown in \Fig{Fig:massdep},
for the $m_q$ dependence of the meson masses from the more realistic
MT ladder-rainbow model~\cite{Maris:1999nt}.  The reason is that the
latter model has a phenomenological infrared strength fitted to
empirical light quark information, $\langle \bar q q \rangle$; it
therefore implicitly includes much of the attraction from vertex
dressing.  The resulting $m_\rho$ is only 5\% off.  The algebraic
model analysis in \Fig{fig:LR_err} suggests that the relative error of
a ladder-rainbow vector mass can only get smaller with increasing
$m_q$; it should drop by a factor of about 5 in going from the $u/d$
quark region to the $c$ quark region, and be about 1\% for $b$ quarks.
That is what we find for the MT model in \Fig{Fig:massdep}; the error
is only 2\% for $c$ quarks and 1\% for $b$ quarks.  Although the
algebraic model is simple and not all details will be realistic, we
expect the qualitative aspects of these observations to be confirmed
whenever a more realistic study becomes feasible.

\subsection{Current conservation}
\begin{figure}[bt]
\centerline{\includegraphics[width=6.4cm]{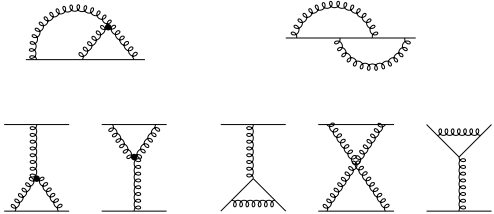}}
\caption{ 
The two leading-order (1-loop) vertex corrections to the rainbow DSE
(top) and the corresponding 5 additions to the ladder BSE kernel
(bottom) to preserve the relevant Ward identities.  Quark and gluon
lines are dressed.
\label{DSEbeyond} }
\end{figure}
If one goes beyond the rainbow-ladder truncation for the propagators,
the meson BSAs and the quark-photon vertex, one has to go beyond
impulse approximation for the meson form factor to ensure current
conservation~\cite{Maris:2000sk}.  For example, following the general
procedure of Ref.~\cite{systematicexp}, one-loop dressing of the
quark-gluon vertex in the DSE requires 5 additions to the ladder BSE
kernel to preserve the relevant WTIs, see Fig.~\ref{DSEbeyond}.

\begin{figure}[bt]
\centerline{\includegraphics[width=6.4cm]{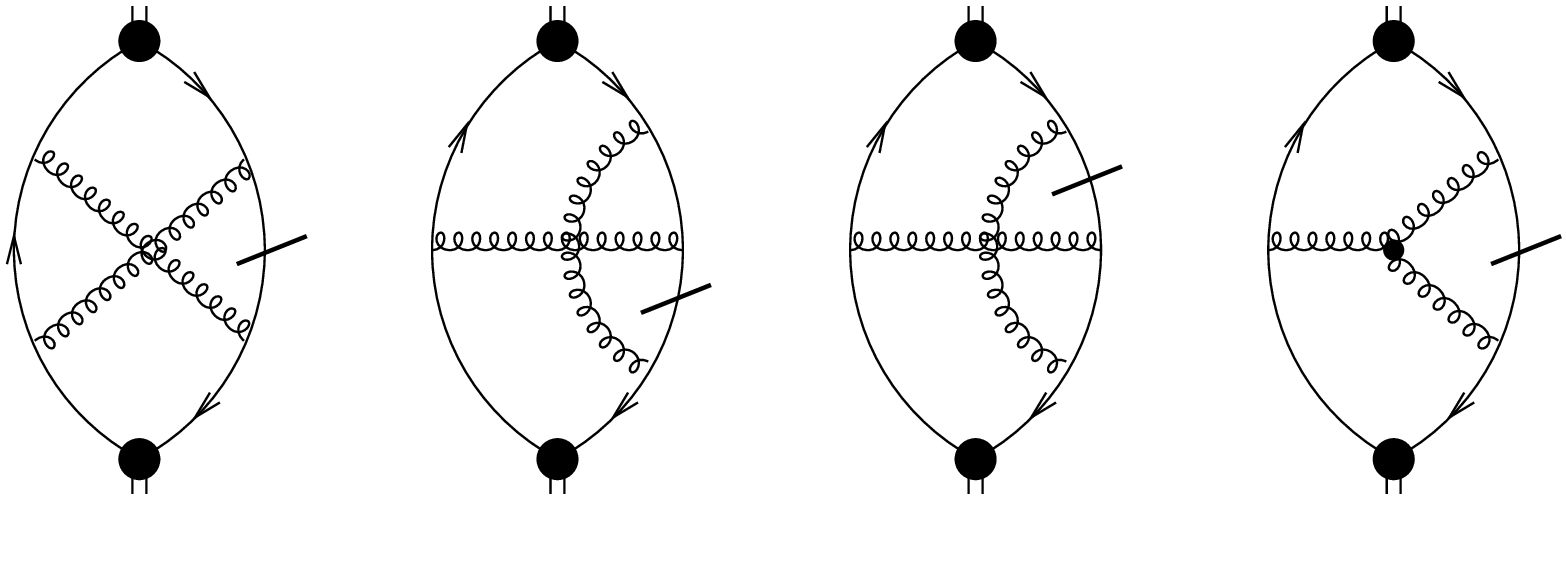}}
\centerline{\includegraphics[width=6.4cm]{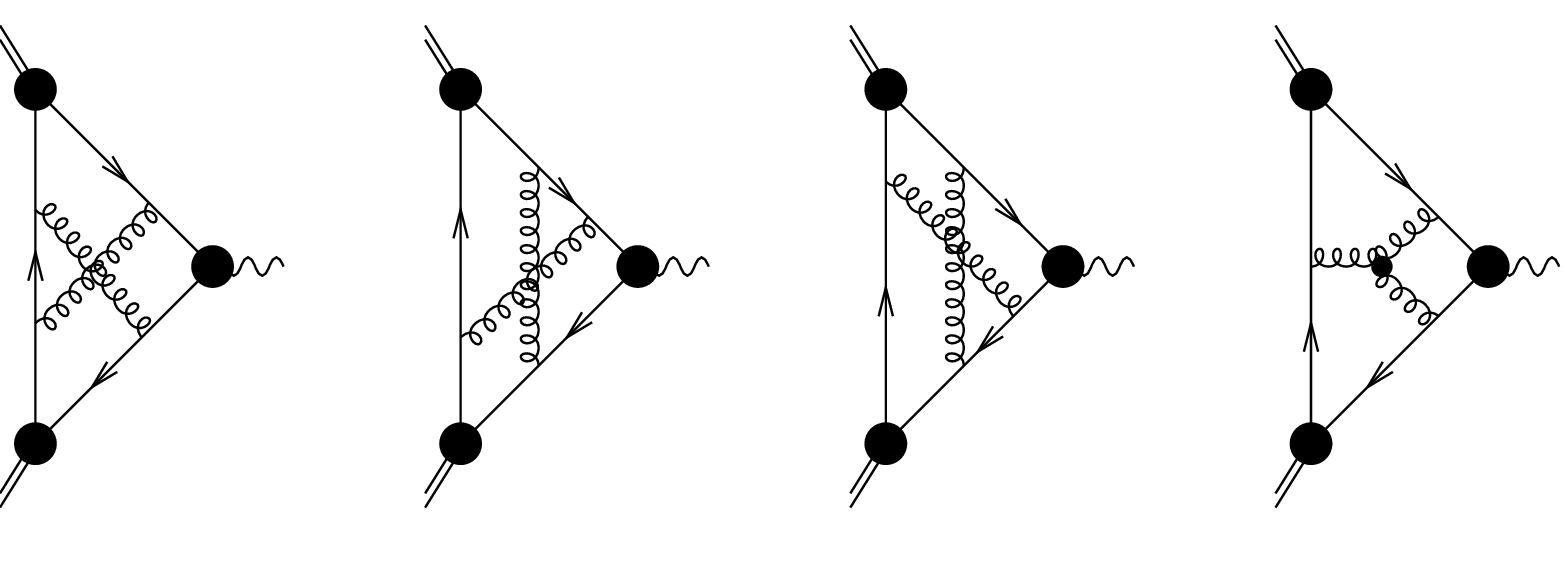}}
\caption{ 
Top: The 4 contributons to the BSA normalization condition from the additions
of Fig.~\protect\ref{DSEbeyond} to the DSE
dynamics.   Bottom: The corresponding 4 corrections to the impulse
approximation to the form factor.   Slashes denote derivatives as
described in the text.  \label{normbeyond} }
\end{figure}

The resulting BSE kernel $K(q_o, q_i; k_o, k_i)$ now becomes dependent
on the total meson momentum $P=q_o-q_i=k_o-k_i$, which means that the
first term in the normalization condition, Eq.(\ref{Eq:BSEnorm}), is
nonzero.  With the choice $k_o = k+P$, $k_i = k$, this introduces the
4 extra terms in the normalization condition depicted in the top row
of Fig.~\ref{normbeyond}.  These 4 additional diagrams are generated
from the BSE kernel in the bottom part of Fig.~\ref{DSEbeyond} by
taking the derivative with respect to the meson momentum $P$, where
$P$ flows through one quark propagator only.  Since a derivative with
respect to $P$ is equivalent to the insertion of a zero-momentum
photon according to the differential Ward identity,
Eq.~(\ref{Eq:diffWI}), it is obvious which diagrams have to be added
to the impulse approximation to maintain current
conservation~\cite{Maris:2000sk}, and they are displayed in the bottom
row of Fig.~\ref{normbeyond}.  The transversality of the photon-meson
coupling is also maintained by this truncation.

We note that this procedure systematically exposes Fock-space
components of the meson that are included in the exact $q\bar q$
scattering kernel of the BSE but which necessarily show up explicitly
as corrections to the impulse approximation.

\section{CONCLUDING REMARKS}

We have shown how the rainbow-ladder truncation of the set of DSEs can
be applied to study pseudoscalar and vector meson properties, from
pions up to $b\bar{b}$ bound states.  This truncation respects the
relevant symmetries and related Ward identities: e.g. the axial-vector
WTI, which ensures the Goldstone nature of the pions in the chiral
limit; we also preserve the correct QCD renormalization properties.
Another advantages of this approach is its manifest Poincar\'e
invariance, which makes the study of electromagnetic form factors
relatively simple.  In addition to the dressed quark propagators and
bound state amplitudes, one needs the quark-photon vertex, which can
be obtained by solving an inhomogeneous BSE~\cite{Maris:1999bh}.  The
rainbow-ladder truncation, in combination with impulse approximation
for the form factor, guarantees current conservation.  Our result for
$F_\pi(Q^2)$ is in remarkable agreement with the data from
TJNAF~\cite{Volmer:2000ek}.

The results for the kaon form factors are also in good agreement with
the data~\cite{Maris:2000sk}, as are the results for the
radiative~\cite{Maris:2000wz,Maris:2002mz} and strong
decays~\cite{Jarecke:2002xd} of the vector mesons, without any
re-adjustments of the model parameters.  Hadronic processes involving
four (or more) external particles can also be described in this
framework, though one has to go beyond impulse approximation.  Our
results for $\pi$-$\pi$ scattering~\cite{pipi} agree with predictions
from chiral symmetry, and this method has been applied to the
anomalous $\gamma \pi\pi\pi$ coupling~\cite{Cotanch:2003xv} as well.
This approach unambiguously incorporates bound state effects in
processes that receive contributions from off-shell intermediate
mesons, such as $\sigma$ and $\rho$ mesons in case of $\pi$-$\pi$
scattering~\cite{pipi}, and VMD effects in the case of electromagnetic
interactions.

Within the same framework, one can consider baryons as bound states of
a quark and a diquark, the latter in a color anti-triplet
configuration.  It has been shown~\cite{Oettel:1998bk} that one can
obtain a reasonable description of the ground state octet and
decouplet baryons using both scalar and axial-vector diquarks.  There
has also been significant progress in understanding the nucleon form
factors using this approach~\cite{baryonemf}.

\section{ACKNOWLEDGEMENTS}

This work was supported by the US Department of Energy, contract
No.~DE-FG02-00ER41135 and by the National Science Foundation, grant
No.~PHY-0301190.  The work benefited from the facilities of the NSF
Terascale Computing System at the Pittsburgh Supercomputing Center.
We thank C.D.~Roberts for very useful discussions and are grateful to
P.~Bowman and J.I.~Skullerud for providing lattice-QCD results.
Appreciation is extended to D.~Leinweber, the Light Cone organizing
committee, and the staff and members of the CSSM, University of
Adelaide for hospitality and support.



\begin{thebibliography}{99}

\bibitem{review}
C.D.~Roberts and A.G.~Williams,
Prog.\ Part.\ Nucl.\ Phys.\  {\bf 33}, 477 (1994)
[arXiv:hep-ph/9403224];
P.C.~Tandy,
Prog.\ Part.\ Nucl.\ Phys.\  {\bf 39}, 117 (1997)
[arXiv:nucl-th/9705018].
C.D.~Roberts and S.M.~Schmidt,
Prog.\ Part.\ Nucl.\ Phys.\  {\bf 45S1}, 1 (2000)
[arXiv:nucl-th/0005064].

\bibitem{Alkofer:2000wg}
R.~Alkofer and L.~von Smekal,
Phys.\ Rept.\  {\bf 353}, 281 (2001)
[arXiv:hep-ph/0007355].

\bibitem{Maris:2003vk}
P.~Maris and C.D.~Roberts,
Int.\ J.\ Mod.\ Phys.\ E {\bf 12}, 297 (2003)
[arXiv:nucl-th/0301049].

\bibitem{systematicexp}
A.~Bender, C.D.~Roberts and L.~Von Smekal,
Phys.\ Lett.\ B {\bf 380}, 7 (1996)
[arXiv:nucl-th/9602012];
A.~Bender, W.~Detmold, C.D.~Roberts and A.W.~Thomas,
Phys.\ Rev.\ C {\bf 65}, 065203 (2002)
[arXiv:nucl-th/0202082].

\bibitem{Maris:1997hd}
P.~Maris, C.D.~Roberts and P.C.~Tandy,
Phys.\ Lett.\ B {\bf 420}, 267 (1998)
[arXiv:nucl-th/9707003].

\bibitem{Maris:1997tm}
P.~Maris and C.D.~Roberts,
Phys.\ Rev.\ C {\bf 56}, 3369 (1997)
[arXiv:nucl-th/9708029].

\bibitem{Roberts:1994hh}
C.D.~Roberts,
Nucl.\ Phys.\ A {\bf 605}, 475 (1996)
[arXiv:hep-ph/9408233].

\bibitem{Maris:1999nt}
P.~Maris and P.C.~Tandy,
Phys.\ Rev.\ C {\bf 60}, 055214 (1999)
[arXiv:nucl-th/9905056].

\bibitem{Bhagwat:2003vw}
M.S.~Bhagwat, M.A.~Pichowsky, C.D.~Roberts and P.C.~Tandy,
Phys.\ Rev.\ C {\bf 68}, 015203 (2003)
[arXiv:nucl-th/0304003].

\bibitem{Fischer:2003rp}
C.S.~Fischer and R.~Alkofer,
Phys.\ Rev.\ D {\bf 67}, 094020 (2003)
[arXiv:hep-ph/0301094].

\bibitem{Politzer:1976tv}
H.D.~Politzer,
Nucl.\ Phys.\ B {\bf 117}, 397 (1976).

\bibitem{Higashijima:1983gx}
K.~Higashijima,
Phys.\ Rev.\ D {\bf 29}, 1228 (1984).

\bibitem{Fomin:1984tv}
P.I.~Fomin, V.P.~Gusynin, V.A.~Miransky and Y.A.~Sitenko,
Riv.\ Nuovo Cim.\  {\bf 6N5}, 1 (1983).

\bibitem{Skullerud:2001aw}
J.~Skullerud, D.B.~Leinweber and A.G.~Williams,
Phys.\ Rev.\ D {\bf 64}, 074508 (2001)
[arXiv:hep-lat/0102013];

\bibitem{Bowman:2002kn}
P.O.~Bowman, U.M.~Heller, D.B.~Leinweber and A.G.~Williams,
Nucl.\ Phys.\ Proc.\ Suppl.\  {\bf 119}, 323 (2003)
[arXiv:hep-lat/0209129].

\bibitem{Bowman:2004xi}
P.O.~Bowman {\it et al.},
Nucl.\ Phys.\ Proc.\ Suppl.\  {\bf 128} (2004) 23
[arXiv:hep-lat/0403002].

\bibitem{Bhagwat:2004hn}
M.S.~Bhagwat {\it et al.},
Phys.\ Rev.\ C {\bf 70}, 035205 (2004)
[arXiv:nucl-th/0403012].

\bibitem{Holl:2004fr}
A.~Holl, A.~Krassnigg and C.D.~Roberts,
Phys.\ Rev.\ C {\bf 70}, 042203 (2004)
[arXiv:nucl-th/0406030].

\bibitem{Ivanov:1997iu}
M.A.~Ivanov, Y.L.~Kalinovsky, P.~Maris and C.D.~Roberts,
Phys.\ Rev.\ C {\bf 57}, 1991 (1998)
[arXiv:nucl-th/9711023].

\bibitem{PDG}
S.~Eidelman {\it et al.}  [Particle Data Group Collaboration],
Phys.\ Lett.\ B {\bf 592}, 1 (2004)

\bibitem{Holl:2005vu}
A.~Holl {\it et al.},
Phys.\ Rev.\ C {\bf 71}, 065204 (2005)
[arXiv:nucl-th/0503043].

\bibitem{Maris:1999bh}
P.~Maris and P.C.~Tandy,
Phys.\ Rev.\ C {\bf 61}, 045202 (2000)
[arXiv:nucl-th/9910033].

\bibitem{Maris:2000sk}
P.~Maris and P.C.~Tandy,
Phys.\ Rev.\ C {\bf 62}, 055204 (2000)
[arXiv:nucl-th/0005015].

\bibitem{Amendolia:1986wj}
S.R.~Amendolia {\it et al.}  [NA7 Collaboration],
Nucl.\ Phys.\ B {\bf 277}, 168 (1986).

\bibitem{Volmer:2000ek}
J.~Volmer {\it et al.}  [The Jefferson Lab F(pi) Collaboration],
Phys.\ Rev.\ Lett.\  {\bf 86}, 1713 (2001)
[arXiv:nucl-ex/0010009].

\bibitem{vanderHeide:2003kh}
J.~van der Heide, J.H.~Koch and E.~Laermann,
Phys.\ Rev.\ D {\bf 69}, 094511 (2004)
[arXiv:hep-lat/0312023].

\bibitem{Bonnet:2004fr}
F.D.R.~Bonnet {\it et al.} [LHP Collaboration],
Phys.\ Rev.\ D {\bf 72}, 054506 (2005)
[arXiv:hep-lat/0411028].

\bibitem{Alkofer:1993gu}
R.~Alkofer, A.~Bender and C.D.~Roberts,
Int.\ J.\ Mod.\ Phys.\ A {\bf 10}, 3319 (1995)
[arXiv:hep-ph/9312243].

\bibitem{Maris:1998hc}
P.~Maris and C.D.~Roberts,
Phys.\ Rev.\ C {\bf 58}, 3659 (1998)
[arXiv:nucl-th/9804062].

\bibitem{Maris:2000wz}
P.~Maris,
Nucl.\ Phys.\ Proc.\ Suppl.\  {\bf 90}, 127 (2000)
[arXiv:nucl-th/0008048].

\bibitem{Maris:2002mz}
P.~Maris and P.C.~Tandy,
Phys.\ Rev.\ C {\bf 65}, 045211 (2002)
[arXiv:nucl-th/0201017].

\bibitem{asympigg}
G.P.~Lepage and S.J.~Brodsky,
Phys.\ Rev.\ D {\bf 22}, 2157 (1980);
M.K.~Chase,
Nucl.\ Phys.\ B {\bf 167}, 125 (1980).

\bibitem{Artuso:2005ym}
M.~Artuso {\it et al.}  [CLEO Collaboration],
arXiv:hep-ex/0508057.

\bibitem{Krassnigg:2004if}
A.~Krassnigg and P.~Maris,
J.\ Phys.\ Conf.\ Ser.\  {\bf 9}, 153 (2005)
[arXiv:nucl-th/0412058].

\bibitem{Alkofer:2002bp}
R.~Alkofer, P.~Watson and H.~Weigel,
Phys.\ Rev.\ D {\bf 65}, 094026 (2002)
[arXiv:hep-ph/0202053].

\bibitem{Watson:2004kd}
P.~Watson, W.~Cassing and P.C.~Tandy,
Few Body Syst.\  {\bf 35}, 129 (2004)
[arXiv:hep-ph/0406340].

\bibitem{Maris:2002yu}
P.~Maris,
Few Body Syst.\  {\bf 32}, 41 (2002)
[arXiv:nucl-th/0204020].

\bibitem{Skullerud:2003qu}
J.I.~Skullerud {\it et al.},
JHEP {\bf 0304}, 047 (2003)
[arXiv:hep-ph/0303176].

\bibitem{Skullerud:2004gp}
J.I.~Skullerud {\it et al.},
Nucl.\ Phys.\ Proc.\ Suppl.\  {\bf 141}, 244 (2005)
[arXiv:hep-lat/0408032].

\bibitem{Bhagwat:2004kj}
M.S.~Bhagwat and P.C.~Tandy,
Phys.\ Rev.\ D {\bf 70}, 094039 (2004)
[arXiv:hep-ph/0407163].

\bibitem{Alkofer:2004it}
R.~Alkofer, C.S.~Fischer and F.J.~Llanes-Estrada,
Phys.\ Lett.\ B {\bf 611}, 279 (2005)
[arXiv:hep-th/0412330].

\bibitem{Leinweber:1998uu}
D.B.~Leinweber, J.I.~Skullerud, A.G.~Williams and C.~Parrinello  
[UKQCD Collaboration],
Phys.\ Rev.\ D {\bf 60}, 094507 (1999)
[Erratum-ibid.\ D {\bf 61}, 079901 (2000)]
[arXiv:hep-lat/9811027].

\bibitem{vonSmekal:1997is}
L.~von Smekal, R.~Alkofer and A.~Hauck,
Phys.\ Rev.\ Lett.\  {\bf 79}, 3591 (1997)
[arXiv:hep-ph/9705242];
ibid,
Annals Phys.\  {\bf 267}, 1 (1998)
[Erratum-ibid.\  {\bf 269}, 182 (1998)]
[arXiv:hep-ph/9707327].

\bibitem{Hawes:1998cw}
F.T.~Hawes, P.~Maris and C.D.~Roberts,
Phys.\ Lett.\ B {\bf 440}, 353 (1998)
[arXiv:nucl-th/9807056].

\bibitem{Davydychev:2000rt}
A.I.~Davydychev, P.~Osland and L.~Saks,
Phys.\ Rev.\ D {\bf 63}, 014022 (2001)
[arXiv:hep-ph/0008171].

\bibitem{Ball:1980ay} 
J.S.~Ball and T.~W.~Chiu,
Phys.\ Rev.\ D {\bf 22}, 2542 (1980).

\bibitem{Munczek:1983dx} 
H.J.~Munczek and A.M.~Nemirovsky, 
Phys.\ Rev.\ {\bf D 28} (1983) 181. 

\bibitem{Jarecke:2002xd}
D.~Jarecke, P.~Maris and P.C.~Tandy,
Phys.\ Rev.\ C {\bf 67}, 035202 (2003)
[arXiv:nucl-th/0208019].

\bibitem{pipi}
P.~Bicudo {\it et al.},
Phys.\ Rev.\ D {\bf 65}, 076008 (2002)
[arXiv:hep-ph/0112015];
S.R.~Cotanch and P.~Maris,
Phys.\ Rev.\ D {\bf 66}, 116010 (2002)
[arXiv:hep-ph/0210151].

\bibitem{Cotanch:2003xv}
S.R.~Cotanch and P.~Maris,
Phys.\ Rev.\ D {\bf 68}, 036006 (2003)
[arXiv:nucl-th/0308008].

\bibitem{Oettel:1998bk}
M.~Oettel, G.~Hellstern, R.~Alkofer and H.~Reinhardt,
Phys.\ Rev.\ C {\bf 58}, 2459 (1998)
[arXiv:nucl-th/9805054].

\bibitem{baryonemf}
R.~Alkofer {\it et al.},
Few Body Syst.\  {\bf 37}, 1 (2005)
[arXiv:nucl-th/0412046];
A.~Holl {\it et al.},
Nucl.\ Phys.\ A {\bf 755}, 298 (2005)
[arXiv:nucl-th/0501033].

\end{thebibliography}
\end{document}